\documentclass[10pt]{article}
\usepackage{bbm}
\usepackage[final]{graphics}
\usepackage{amsmath}
\usepackage{amsfonts,amsbsy}
\usepackage{amssymb}

\def\empile#1\over#2{\mathrel{\mathop{\kern 0pt#1}\limits_{#2}}}

\newcommand{\slv}{\raise.15ex\hbox{$/$}\kern-.53em\hbox{$v$}}
\newcommand{\slF}{\raise.15ex\hbox{$/$}\kern-.53em\hbox{$F$}}
\newcommand{\slL}{\raise.15ex\hbox{$/$}\kern-.53em\hbox{$L$}}
\newcommand{\slP}{\raise.15ex\hbox{$/$}\kern-.53em\hbox{$P$}}
\newcommand{\slp}{\raise.15ex\hbox{$/$}\kern-.53em\hbox{$p$}}
\newcommand{\slq}{\raise.15ex\hbox{$/$}\kern-.53em\hbox{$q$}}
\newcommand{\slR}{\raise.15ex\hbox{$/$}\kern-.53em\hbox{$R$}}
\newcommand{\slQ}{\raise.15ex\hbox{$/$}\kern-.53em\hbox{$Q$}}
\newcommand{\slK}{\raise.15ex\hbox{$/$}\kern-.53em\hbox{$K$}}
\newcommand{\slk}{\raise.15ex\hbox{$/$}\kern-.53em\hbox{$k$}}
\newcommand{\slD}{\raise.15ex\hbox{$/$}\kern-.73em\hbox{$D$}}
\newcommand{\slC}{\raise.15ex\hbox{$/$}\kern-.53em\hbox{$C$}}
\newcommand{\slA}{\raise.15ex\hbox{$/$}\kern-.53em\hbox{$A$}}
\newcommand{\slSigma}{\raise.15ex\hbox{$/$}\kern-.53em\hbox{$\Sigma$}}
\newcommand{\slpartial}{\raise.15ex\hbox{$/$}\kern-.53em\hbox{$\partial$}}
\newcommand{\slcalP}{\raise.15ex\hbox{$/$}\kern-.63em\hbox{$\cal P$}}

\def\l{{\boldsymbol l}}
\def\k{{\boldsymbol k}}

\def\x{{\boldsymbol x}}
\def\y{{\boldsymbol y}}

\def\v{{\boldsymbol v}}

\def\u{{\boldsymbol u}}

\catcode`\@=11

\newcount\@tempcntc
\def\@citex[#1]#2{\if@filesw\immediate\write\@auxout{\string\citation{#2}}\fi
  \@tempcnta\z@\@tempcntb\m@ne\def\@citea{}\@cite{%
        \@for\@citeb:=#2\do%
    {\@ifundefined{b@\@citeb}%
        {\@citeo\@tempcntb\m@ne\@citea%
                \def\@citea{,\penalty\@m\ }{\bf ?}\@warning%
                {Citation `\@citeb' on page \thepage \space undefined}}%
        {\setbox\z@\hbox{\global\@tempcntc0\csname b@\@citeb\endcsname\relax}
     \ifnum\@tempcntc=\z@ \@citeo\@tempcntb\m@ne%
       \@citea\def\@citea{,\penalty\@m}%
       \hbox{\csname b@\@citeb\endcsname}%
     \else%
      \advance\@tempcntb\@ne%
      \ifnum\@tempcntb=\@tempcntc%
      \else\advance\@tempcntb\m@ne\@citeo%
      \@tempcnta\@tempcntc\@tempcntb\@tempcntc\fi\fi}}\@citeo}{#1}}%

\def\@citeo{\ifnum\@tempcnta>\@tempcntb\else\@citea
  \def\@citea{,\penalty\@m}%
  \ifnum\@tempcnta=\@tempcntb\the\@tempcnta\else
   {\advance\@tempcnta\@ne\ifnum\@tempcnta=\@tempcntb \else
\def\@citea{--}\fi
    \advance\@tempcnta\m@ne\the\@tempcnta\@citea\the\@tempcntb}\fi\fi}

\catcode`\@=12

\begin{document}

\title{\bf %
Role of quantum fluctuations\\ 
in a system with strong fields:\\
Onset of hydrodynamical flow}
\author{Kevin Dusling$^{(1)}$, Thomas Epelbaum$^{(2)}$,\\ Fran\c cois Gelis$^{(2)}$, Raju Venugopalan$^{(1)}$}
\maketitle
\begin{center}
\begin{enumerate}
\item  Physics Department\\
  Brookhaven National Laboratory\\
  Upton, NY-11973, USA
\item Institut de Physique Th\'eorique (URA 2306 du CNRS)\\
  CEA/DSM/Saclay\\
  91191, Gif-sur-Yvette Cedex, France
\end{enumerate}
\end{center}

\maketitle

\begin{abstract}
  Quantum fluctuations are believed to play an important role in the
  thermalization of classical fields in inflationary cosmology but
  their relevance for isotropization/thermalization of the classical
  fields produced in heavy ion collisions is not completely
  understood.  We consider a scalar $\phi^4$ toy model coupled to a
  strong external source, like in the Color Glass Condensate
  description of the early time dynamics of ultrarelativistic heavy
  ion collisions. The leading order classical evolution of the scalar
  fields is significantly modified by the rapid growth of
  time-dependent quantum fluctuations, necessitating an all order
  resummation of such ``secular'' terms. We show that the resummed
  expressions cause the system to evolve in accordance with ideal
  hydrodynamics.  We comment briefly on the thermalization of our
  quantum system and the extension of our results to a gauge theory.
\end{abstract}

\section{Introduction}
A remarkable outcome of the heavy ion experiments at
RHIC~\cite{Adamsa3,Adcoxa1,Arsena2,Backa2} is the large elliptic flow
observed in the collisions. Phenomenological hydrodynamical models
that fit the RHIC data appear to require that the quark gluon matter
has a very small value for the dimensionless ratio of the viscosity to
the entropy density~\cite{Romat1}. This ratio $\eta/s$, a measure of
the ``perfect fluidity'' of the system, is estimated to be $\lesssim
5/4\pi$~\cite{Teane1}, where $\eta/s = 1/4\pi$ is a conjectured
universal lower bound~\cite{PolicSS1,PolicSS2}. Its was shown
recently~\cite{HiranHKLN1,LappiV1,DrescDHN1} that the degree of
perfectness of the quark-gluon fluid produced at RHIC is sensitive to
details of the initial spatial distribution of the produced matter at
the onset of hydrodynamic flow.

An important feature of the hydrodynamic models is that they require
very early thermalization after the collision. Estimates for the
thermalization time, which range from $\tau_{\rm relax}\sim 0.6 -1$
fm~\cite{SongH1,LuzumR1,DusliMT1}, are difficult to reconcile with a
simple picture of thermalization arising from the rapid scattering of
quasi-particles at rates greater than the expansion rate of the fluid.
The uncertainty principle tells us that for $\tau_{\rm relax} \leq 1$
fm, modes with momenta $\sim 200$ MeV are not even on-shell, let alone
amenable to being described as quasi-particles undergoing scattering.
While a quasi-particle description is not essential to thermalization,
it is the simplest one, and other realizations are more complicated.
With regard to the issue of flow however, it is sufficient to note
that one requires primarily that matter be isotropic and (nearly)
conformal to obtain a closed form expression for the hydrodynamic
equations~\cite{ArnolLMY1}.

How isotropization and (subsequently) thermalization is achieved in
heavy ion collisions is an outstanding problem which requires that the
problem be considered {\it ab initio}. What this means it that one
needs to understand and compute the properties of the relevant degrees
of freedom in the nuclear wavefunctions and how these degrees of
freedom decohere in a collision to produce quark-gluon matter. An {\it
  ab initio} approach to the problem can be formulated within the
framework of the Color Glass Condensate (CGC) effective field theory,
which describes the relevant degrees of freedom in the nuclei as
dynamical classical fields coupled to static color
sources~\cite{IancuV1,IancuLM3,GelisIJV1}. The computational power of
this approach is a consequence of the dynamical generation of a
semi-hard scale, the saturation scale~\cite{GriboLR1,MuellQ1}, which
allows a weak coupling treatment of the relevant degrees of
freedom~\cite{McLerV1,McLerV2,McLerV3} in the high energy
nuclear wavefunctions.

There has been significant progress recently in applying the CGC
effective field theory to studying the early time behavior of the
matter produced in heavy ion collisions. Inclusive quantities such as
the pressure and the energy density in this matter (called the
Glasma~\cite{LappiM1}) can be written as expressions that factorize
the universal properties of the nuclear wavefunctions (measurable for
instance in proton-nucleus or electron-nucleus collisions) from the
detailed dynamics of the matter in
collision~\cite{GelisLV3,GelisLV4,GelisLV5}. Key to this approach are
the quantum fluctuations around the classical fields in the
wavefunctions and in the collision. Quantum fluctuations that are
invariant under boosts can be isolated in universal functionals that
evolve with energy. There are however also quantum fluctuations that
are not boost invariant which are generated during the collision.
These quantum fluctuations can grow rapidly and therefore play a
significant role in the subsequent temporal evolution of the Glasma.

The problem of how to treat these so-called ``secular divergences'' of
perturbative series is very general and occurs in a wide variety of
dynamical systems~\cite{Golde1}. In particular, the role of time
dependent quantum fluctuations in heavy ion collisions bears a strong
analogy to their role in the evolution of the early
universe~\cite{AllahBCM1}. In the latter case, quantum fluctuations
around a rapidly decaying classical field, the inflaton, are enhanced
due to parametric resonance, and it is conjectured that this dynamics
termed ``preheating''~\cite{GreenKLS1} may lead to turbulent
thermalization~\cite{MichaT1} in the early universe.

It is therefore very important to understand the precise role of these
quantum fluctuations in heavy ion collisions to determine whether they
play an analogous role to that in the early universe in the
isotropization/thermalization of the system. Their computation in a
gauge theory is quite involved so for simplicity, we shall in this
paper first attack this problem in a scalar $\phi^4$ field theory.
Like QCD, the coupling is dimensionless in this theory and the fields
are self interacting. In addition, we choose initial conditions for
our study that are similar to those in the CGC treatment of heavy ion
collisions. It must be said at the outset that there are important
differences between the two theories and there is no {\it a priori}
guarantee that the lessons learnt in one case will translate
automatically to the other.

The CGC initial conditions, for weak couplings $g\ll 1$, specifically
lead to a power counting scheme where the leading contribution to
inclusive quantities is the classical contribution of order ${\cal
  O}(1/g^2)$.  Quantum corrections begin at ${\cal O}(1)$ and their
contribution can be expressed as real-time partial differential
equations for small fluctuations in the classical background, with
purely retarded initial conditions. We will show that there are modes
of the small fluctuation field that grow very rapidly and can become
as large as the classical field on time scales of interest in the
problem. We observe that there are two sorts of rapidly growing modes
of the fluctuation field. One are modes that enjoy parametric
resonance and grow exponentially. These modes are however localized in
a rather narrow resonance band. The zero mode and low lying modes grow
linearly and can also influence the temporal evolution of the system.
Both sorts of ``secular'' terms can be isolated and resummed to all
orders in perturbation theory. The resulting expressions are stable
and can be expressed as an ensemble average over a spectrum of quantum
fluctuations convolved with the {\it leading order} inclusive quantity
which, for a particular fluctuation field, is a functional of the
classical field shifted by that quantum fluctuation. We note that a
similar observation was made previously in the context of inflationary
cosmology~\cite{Son1,PolarS1,KhlebT1}.

The fact that one can express resummed expressions for the pressure
and energy density as ensemble averages over quantum fluctuations has
profound consequences.  Without resummation, the relation between the
energy density and the pressure is not single valued. For the resummed
expressions, while the relation between the pressure and energy
density is not single valued at early times, it becomes so after a
finite evolution time. This development of an ``equation of state''
therefore allows one to write the conservation equation for the
resummed energy momentum tensor $T^{\mu\nu}$ as a closed form set of
equations, which are the equations of ideal hydrodynamics. This of
course suggests that the system behaves as a perfect fluid. If the
considerations in our paper can be applied to a gauge theory, the
result would have significant ramifications for the interpretation of
the heavy ion experiments and the extraction of $\eta/s$ in
hydrodynamical models.

The evolution of the system towards the equation of state
characteristic of hydrodynamic flow can be interpreted as arising from
a phase decoherence of the different classical trajectories of the
energy momentum tensor for different initial conditions given by the
ensemble of quantum fluctuations. For a scalar $\phi^4$ theory, the
frequency of the periodic classical trajectories is proportional to
the amplitude. Therefore, for different initial values of the
amplitude, the different trajectories are phase shifted. The ensuing
cancellations between trajectories results in the single valued
relation between the pressure and the energy density. While it appears
that decoherence can arise from the zero mode and near lying modes
alone, the inclusion of the resonant band significantly alters the
decoherence of the system. Similar behavior has been seen in models of
reheating after inflation \cite{ProkoR1,Frolo1,FeldeT1}.  In
particular, one sees that quantum de-coherence of the inflaton field
leads to a transition from a dust--like equation of state to a
radiation dominated era.

It is interesting to ask whether the decoherence and concomitant
fluidity observed in our numerical simulations implies thermalization
of the system. We first investigate the behavior of the ensemble of
initial conditions in the {\it Poincar{\'e}} phase plane for the toy
case of uniform background field and fluctuations. One sees that the
initially localized trajectories spread around a close loop filling
the phase-space as one would expect for the phase-space density of a
micro-canonical ensemble. For the toy example considered, the ensemble
average of the trace of the energy momentum tensor can be expressed at
large times as the time average along a single trajectory in the {\it
  Poincar{\'e}} phase plane. For the scale invariant $\phi^4$ theory,
this average is zero with the consequence that the energy momentum
tensor becomes traceless resulting in a single valued relation between
the energy density and the pressure.

Going beyond the toy example, for the general case of spatially
non-uniform fluctuations, there is no easy way to visualize trajectories on the {\it
  Poincar{\'e}} phase plane because the system is infinite dimensional. However, because the numerical problem
is formulated on a lattice, one can look at a small sub-system on this
lattice and study its event-by-event energy fluctuations. Starting
from a Gaussian initial distribution, we see that the distribution
converges to an exponential form. One can also study the moments of
the energy distribution; these again demonstrate a rapid change from
initial transient values to stationary values. While the behavior is
close to those expected from a canonical thermal ensemble, it is
premature from our present studies to make definitive conclusions.
This will require a careful study of the effects of varying the
coupling and volume effects and will be left to a future study.

We note however that the formalism developed in our paper is well
suited to the study of thermalization of quantum systems\footnote{We
  thank Giorgio Torrieri for bringing to our attention Berry's
  conjecture and the accompanying literature on eigenstate
  thermalization.}. It has been argued
previously~\cite{Deuts1,Sredn1,Jarzy1,RigolDO1} that quantum systems
will thermalize if they satisfy Berry's conjecture~\cite{Berry1}. This
conjecture states that the high lying quantum eigenstates of a system
whose classical behavior is chaotic and ergodic have a wavefunction
that behaves as a linear superposition of plane waves whose
coefficients are Gaussian random variables.  When an inclusive
measurement is performed on such an eigenstate, one obtains results
that agree with the predictions of the micro-canonical equilibrium
ensemble, a property that has been dubbed ``eigenstate
thermalization'' in~\cite{Sredn1}. If the state at $t=0$ is a coherent
superposition of such eigenstates, the micro-canonical predictions
become valid only after the states in the superposition have
sufficiently decohered--thus for quantum systems where Berry's
conjecture apply, thermalization appears to be a consequence of
decoherence. The ensemble of quantum fluctuations included via the
resummation we develop in the section 2.6 leads to fields that have
precisely this behavior (see
eqs.~(\ref{eq:berry-1}-\ref{eq:berry-2})).  Our interest ultimately is
in QCD, where the classical behavior of the system is believed to be
chaotic~\cite{BiroGMT1,HeinzHLMM1,KunihMOST1}.  Because much of our
formalism can be extended to gauge theories, we anticipate that a
first principles treatment of thermalization is feasible.

This paper is organized as follows. In section 2, we introduce the
model scalar theory and the CGC-like initial conditions for its
temporal evolution. We then discuss the computation of $T^{\mu\nu}$ at
leading and next-to-leading order. The problem of secular divergences
is noted, and a stable resummation procedure is developed. A
simplified toy model is considered in section 3, wherein only
spatially uniform fluctuations are considered. The behavior of the
resummed pressure and energy density and their relaxation to an
equation of state is studied. These results are interpreted and
understood as a consequence of the decoherence of the system which
allows one to equate ensemble averages to a temporal average over
individual classical trajectories. For the longitudinally expanding
case, temporal evolution in the toy model displays the behavior of a
fluid undergoing ideal hydrodynamic flow.  The full quantum field
theory is considered in section 4, where we compute {\it ab initio}
the spectrum of fluctuations. The full theory displays the same
essential features as the toy model studied in section 3, albeit the
interplay of linearly growing low lying momentum modes and the
resonant modes leads to a more complex temporal evolution. In this
section, we also investigate the dependence of the relaxation time on
the strength of the coupling constant. Then, we study the energy
distribution in a small subsystem, and its time evolution.  We
conclude with a brief outlook.  Much of the details of the computation
are given in appendices.  In appendix A, we discuss the numerical
solution of the scalar field model, including the lattice
discretization, the computation of the quantum fluctuation spectrum
and the sensitivity of the results to the ultraviolet cut-off. The
stability analysis of linearized perturbations to the classical field
is considered in Appendix B. The resonance band is identified and the
Lyapunov exponents are computed explicitly. We also discuss the
relationship between decoherence and linear instabilities.

\section{Temporal evolution of $T^{\mu\nu}$}

In this section, we will consider a scalar field toy model whose
behavior mimics key features of the Glasma~\cite{LappiM1} description
of the early behavior of the quark-gluon matter produced in high
energy heavy ion collisions. In the CGC framework, strong color fields
are present in the initial conditions for the evolution of the Glasma.
In this situation, the leading order contribution is given by
classical fields, with higher order corrections coming from the {\sl
  apparently} sub-leading quantum fluctuations. We consider the
stress-energy tensor in this scalar field model and discuss its
temporal evolution at leading (LO) and next-to-leading orders (NLO). We
show explicitly that there are contributions at NLO that can grow
with time and become larger than the LO terms. We end this section by
describing how these ``secular'' terms can be resummed and the results
expressed in terms of an average over a Gaussian ensemble of classical
fields.

\subsection{Scalar model with CGC-like initial conditions}
Our CGC inspired scalar model has the Lagrangean
\begin{equation}
{\cal L}\equiv \frac{1}{2}(\partial_\mu\phi)(\partial^\mu\phi)-\frac{g^2}{4!}\phi^4+J\phi\; ,
\label{eq:L}
\end{equation}
where $J$ is an external source. In the CGC framework, the source $J$
coupled to the gauge fields represents the color charge current
carried by the two colliding heavy ions. The current is zero at
positive proper time, corresponding to times after the collision has
taken place. We emulate this feature of the CGC in a simpler
coordinate system by taking the source $J$ to be nonzero only for
Cartesian time $x^0<0$, and parameterize it as\footnote{In the
  numerical implementation of the model, the time dependent prefactor
  is constrained to vanish when $x^0\to-\infty$ to ensure a free
  theory in the remote past.}
\begin{equation}
J(x)\sim \theta(-x^0)\frac{Q^3}{g}\; .
\label{eq:J}
\end{equation}
At $x^0>0$, where $J$ is zero, the fields evolve solely via their
self-interactions, in an analogous fashion to the non-Abelian color
fields produced in the collision of two hadrons or nuclei.

In eq.~(\ref{eq:J}), we incorporated two additional features of the
CGC. The first feature corresponds to a strong external current $J$,
which follows from the power of the inverse coupling when $g\ll 1$;
weak coupling is essential to motivate an expansion in powers of
$g^2$.  The other feature of the CGC that is emulated is that the
dimensionful parameter $Q$ in eq.~(\ref{eq:J}) plays a role analogous
to that of the saturation scale~\cite{GriboLR1,MuellQ1}, in the sense
that non-linear interactions are sizeable for modes $|\k|\lesssim Q$.

Note that a scalar field theory with a $\phi^4$ coupling in four
space-time dimensions is scale invariant at the classical level--the
coupling constant $g$ is dimensionless in the theory. In our model,
this scale invariance is broken by the coupling of the scalar field to
the external source $J$ containing the dimensionful scale $Q$. We may
therefore anticipate that all physical quantities are simply expressed
by the appropriate power of $Q$ times a prefactor that depends on $g$.

\subsection{$T^{\mu\nu}$ at leading order}
Because the source $J$ contains a power of the inverse coupling, the
power counting for Feynman diagrams indicates that the order of
magnitude of a given graph depends only on its number of external
lines and number of loops, but not on the number of sources $J$
attached to the graph~\cite{GelisV2,GelisV3}.  For the
energy-momentum tensor of the theory, the various contributions can be
organized in a series in powers of $g^2$ as
\begin{equation}
T^{\mu\nu}
=
\frac{Q^4}{g^2}
\Big[
c_0+c_1 g^2 +c_2 g^4+\cdots
\Big]\; .
\end{equation}
In this expansion, the coefficients $c_0, c_1, c_2,\cdots$ are
themselves infinite series in the combination $gJ$ corresponding to an
infinite set of Feynman diagrams. This combination is parametrically
independent of $g$ because $J\sim g^{-1}$. More precisely, $c_0$
contains only tree diagrams, $c_1$ 1-loop diagrams, $c_2$ 2-loop
diagrams, and so on.

In our model, the leading order (tree level) contribution to the
energy-momentum tensor can be expressed solely in terms of a classical
solution $\varphi$ of the field equation of motion~\cite{GelisV2}.
Namely, one has
\begin{equation}
T^{\mu\nu}_{_{\rm LO}}(x)
=
c_0\frac{Q^4}{g^2}
=
\partial^\mu\varphi\partial^\nu\varphi
-
g^{\mu\nu}\,\Big[\frac{1}{2}(\partial_\alpha\varphi)^2-\frac{g^2}{4!}\varphi^4\Big]\; ,
\end{equation}
where
\begin{eqnarray}
&&\square \varphi +\frac{g^2}{3!}\varphi^3=J\; ,
\nonumber\\
&&\lim_{x^0\to -\infty}\varphi(x^0,\x)=0\; .
\label{eq:EOM-LO}
\end{eqnarray}
Clearly, due to the non-linear term in the equation of motion, the
solution $\varphi$ (and hence the coefficient $c_0$) depends on $gJ$
to all orders, as stated previously. This LO energy momentum tensor
is conserved\footnote{Strictly speaking, this is true only at $x^0>0$.
  At negative times, some energy is injected into the system by the
  external source $J$.},
\begin{equation}
\partial_\mu T^{\mu\nu}_{_{\rm LO}}=0\; .
\end{equation}

\begin{figure}[htbp]
\begin{center}
\resizebox*{8cm}{!}{\rotatebox{-90}{\includegraphics{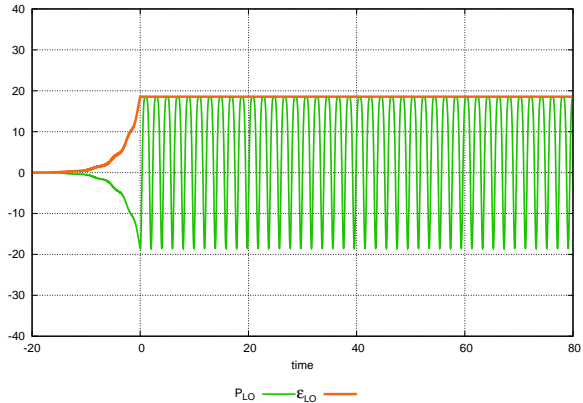}}}
\end{center}
\caption{\label{fig:LO} Components of $T^{\mu\nu}_{_{\rm LO}}$ for a
  spatially uniform external source. To perform this calculation, we
  took in eq.~(\ref{eq:EOM-LO}) a source
  $J=g^{-1}Q^3\theta(-x^0)e^{bQx^0}$ (with $g=1, b=0.1$ and $Q=2.5$),
  that vanishes adiabatically in the remote past.}
\end{figure}
If the source $J$ is taken to be spatially homogeneous, then the
energy-momentum tensor evaluated at leading order has the simple
form
\begin{equation}
T^{\mu\nu}_{_{\rm LO}}(x)
=
\begin{pmatrix}
\epsilon_{_{\rm LO}} & 0& 0& 0\\
0& p_{_{\rm LO}}& 0& 0\\
0& 0& p_{_{\rm LO}}& 0\\
0& 0& 0& p_{_{\rm LO}}\\
\end{pmatrix}\; ,
\end{equation}
with the leading order energy density and pressure given by
\begin{eqnarray}
\epsilon_{_{\rm LO}}&=& \frac{1}{2}\dot\varphi^2 + \frac{g^2}{4!}\varphi^4\nonumber\\
p_{_{\rm LO}}&=& \frac{1}{2}\dot\varphi^2 - \frac{g^2}{4!}\varphi^4\; .
\label{eq:edens-LO}
\end{eqnarray}
One can easily check that the energy density $\epsilon_{_{\rm LO}}$ is
constant in time at $x^0>0$ (after the external source $J$ has been
switched off), while the pressure $p_{_{\rm LO}}$ is a periodic
function of time at $x^0>0$, as illustrated in the figure
\ref{fig:LO}.  From the numerical computation, it is clear that at
this order of the calculation of $\epsilon_{_{\rm LO}}$ and $p_{_{\rm
    LO}}$, one does not have a well defined (single valued)
relationship $\epsilon_{_{\rm LO}} = f(p_{_{\rm LO}})$. In other
words, {\sl there is no equation of state at leading order in $g^2$}.
This might appear problematic at the outset because one might expect
that the scale invariance of the theory would require the energy
momentum tensor to be traceless. As discussed further in section
\ref{sec:qavg}, this is not so for the case of a scalar theory.

\subsection{$T^{\mu\nu}$ at next to leading order}
\label{sec:NLO}
At next-to-leading order, the energy momentum tensor can be written as
\begin{eqnarray}
T^{\mu\nu}_{_{\rm NLO}}
&=&c_1 Q^4=
\partial^\mu\varphi\partial^\nu\beta+\partial^\mu\beta\partial^\nu\varphi
-g^{\mu\nu}\Big[\partial_\alpha\beta\partial^\alpha\varphi
-\beta V^\prime(\varphi)
\Big]
+\nonumber\\
&&
+
\!\int\!\frac{d^3\k}{(2\pi)^3 2 k}
\Big[
\partial^\mu a_{-\k}\partial^\nu a_{+\k}
\!-\!
\frac{g^{\mu\nu}}{2}\Big(\partial_\alpha a_{-\k}\partial^\alpha a_{+\k}-V^{\prime\prime}(\varphi)a_{-\k}a_{+\k}\Big)
\Big]\; ,
\nonumber\\
&&
\label{eq:NLO}
\end{eqnarray}
where for brevity we use the notation $V(\varphi)\equiv
g^2\varphi^4/4!$ with each prime denoting a derivative with respect to
$\varphi$.  In this formula, $\beta$ and $a_{\pm\k}$ are small field
perturbations, that are defined by the following equations:
\begin{eqnarray}
&&\Big[\square+V^{\prime\prime}(\varphi)\Big]a_{\pm\k}=0\nonumber\\
&&\lim_{x^0\to-\infty}a_{\pm\k}(x)=e^{\pm ik\cdot x}\; ,\nonumber\\
&&\Big[\square+V^{\prime\prime}(\varphi)\Big]\beta = 
-\frac{1}{2}V^{\prime\prime\prime}(\varphi)
\int\frac{d^3\k}{(2\pi)^3 2k}\;a_{-\k}a_{+\k}\nonumber\\
&&\lim_{x^0\to-\infty}\beta(x)=0\; .
\label{eq:fluctuations}
\end{eqnarray}
Because the classical field $\varphi$ is
spatially homogeneous in the toy model considered here, the equation of motion for $a_{\pm\k}$
simplifies to
\begin{equation}
\ddot{a}_{\pm\k}+(\k^2+V^{\prime\prime}(\varphi))a_{\pm\k}=0\; ,
\end{equation}
and the field fluctuation $\beta$ depends only on time. 

After some algebra, it is easy to check that the energy-momentum
tensor is also conserved at NLO\footnote{This result should be
  self-evident because the conservation equation $\partial_\mu
  T^{\mu\nu}=0$ is linear in the components of $T^{\mu\nu}$.
  Therefore, it does not mix the different $g^2$ orders, requiring the
  conservation equation to be satisfied for each order in $g^2$.}  for
$x^0>0$,
\begin{equation}
\partial_\mu T^{\mu\nu}_{_{\rm NLO}}=0\; .
\label{eq:cons-NLO}
\end{equation}
The $00$ component of $T^{\mu\nu}_{_{\rm NLO}}$ in eq.~(\ref{eq:NLO})
gives us the energy density at NLO,
\begin{equation}
\epsilon_{_{\rm NLO}}
=
\dot\beta\dot\varphi+\beta V^\prime(\varphi)
+\frac{1}{2}
\int\frac{d^3\k}{(2\pi)^3 2k}\;\Big[
\dot{a}_{-\k}\dot{a}_{+\k}+(\k^2+V^{\prime\prime}(\varphi))a_{-\k}a_{+\k}
\Big]\; .
\label{eq:edens_NLO}
\end{equation}
Given eqs.~(\ref{eq:fluctuations}), it is straightforward to verify
that this correction is also constant in time, $\dot\epsilon_{_{\rm
    NLO}}=0$, in agreement with eq.~(\ref{eq:cons-NLO}).  The $11$
component of eq.~(\ref{eq:NLO}) --the NLO pressure in the $x$
direction-- reads
\begin{equation}
p_{_{\rm NLO}}
=
\dot\beta\dot\varphi -\beta V^\prime(\varphi)
+\frac{1}{2}
\int\frac{d^3\k}{(2\pi)^3 2k}\;\Big[
\dot{a}_{-\k}\dot{a}_{+\k}-(\k^2-2k_x^2+V^{\prime\prime}(\varphi))a_{-\k}a_{+\k}
\Big]\; .
\label{eq:pressure_NLO}
\end{equation}
Note that although the integrand is not rotationally invariant, the
result of the $\k$ integration is symmetric and the NLO pressures are
the same in all directions.

\begin{figure}[htbp]
\begin{center}
\resizebox*{8cm}{!}{\rotatebox{-90}{\includegraphics{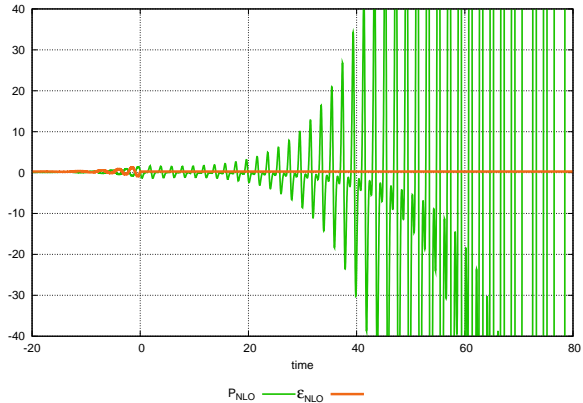}}}
\end{center}
\caption{\label{fig:NLO} Components of $T^{\mu\nu}_{_{\rm NLO}}$ for a
  spatially uniform external source. This calculation was performed 
  for $g=1$.}
\end{figure}

We evaluated numerically $\epsilon_{_{\rm NLO}}$ and $p_{_{\rm NLO}}$
for a coupling constant $g=1$, by first solving
eqs.~(\ref{eq:fluctuations}) for $\beta$ and for the $a_\k$'s (for a
discretized set of $\k$'s). The results of this calculation are shown
in the figure \ref{fig:NLO}.
From this evaluation, we see that the energy density at NLO is
constant at $x^0>0$, as we expected\footnote{This time independence
  can be seen as a test of the accuracy of the numerical calculation,
  because it results from a cancellation between several terms that
  grow with time.}.  We also notice that for $g=1$, the NLO correction
to the energy density is very small, of the order of $1.4\%$ of the LO
result\footnote{Indeed, since there is a prefactor $1/4!$ in our
  definition of the interaction potential, $g=1$ corresponds to fairly
  weak interactions.}.  Thus, we conclude from this that for such a
value of the coupling, we have a well behaved perturbative expansion
for $\epsilon$. The NLO pressure however behaves quite differently.
Not only it is varying in time (hence no equation of state at NLO),
but it also has oscillations whose amplitude grows exponentially at
large $x^0$.  Therefore, the NLO correction to the pressure eventually
becomes larger than the LO contribution, and the perturbative
expansion for the pressure in powers of $g^2$ breaks down\footnote{A
  similar behavior was observed in a different context in
  \cite{BoyanVHLS1}.}.  Also noteworthy is the fact that at $x^0=0$,
$p_{_{\rm NLO}}$ is still a small correction to $p_{_{\rm LO}}$; it
only becomes large at later times.

\subsection{Interpretation of the NLO result}
The secular divergence of the pressure at NLO can be understood as a
consequence of the unstable behavior of $a_{\pm\k}(x)$ for some values
of $\k$. The stability analysis of small quantum fluctuations in
$\phi^4$ field theory is performed in appendix \ref{app:stability}.
From this study, one obtains the following results:
\begin{itemize}
\item[{\bf i.}] There is a range in $|\k|$ where the $a_{\pm\k}$'s diverge
  exponentially in time, due to the phenomenon of parametric resonance.
\item[{\bf ii.}] The zero mode $\k=0$ fluctuation, $a_0$, diverges
  linearly in time, a phenomenon closely related to the fact that the
  oscillation frequency in a non-harmonic potential depends on the
  amplitude of the oscillations.
\end{itemize}
In addition, one observes numerically that fluctuation modes in the
vicinity of $\k=0$, albeit not mathematically unstable, can attain
quite large values. (They appear to grow linearly for some time before
decreasing in value.)

\begin{figure}[htbp]
\begin{center}
\resizebox*{8cm}{!}{\rotatebox{-90}{\includegraphics{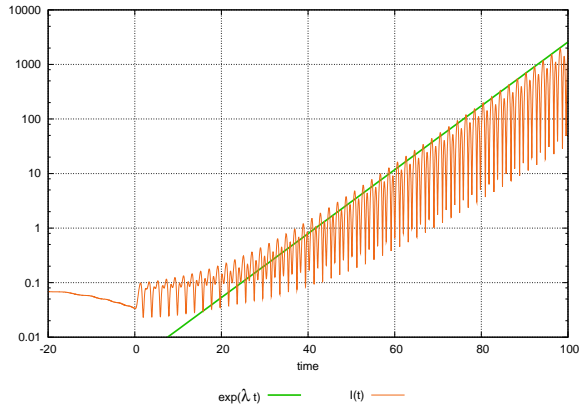}}}
\end{center}
\caption{\label{fig:I} Numerical evaluation of the integral defined in
  eq.~(\ref{eq:I}). The line denotes an exponential fit to the envelope.}
\end{figure}

Because of the existence of modes that grow in time, integrals such as
\begin{equation}
I(x^0)\equiv\int \frac{d^3\k}{(2\pi)^3 2k}\; a_{-\k}(x)a_{+\k}(x)\; ,
\label{eq:I}
\end{equation}
that appear in the components of $T^{\mu\nu}_{_{\rm NLO}}$ (see
eqs.~(\ref{eq:edens_NLO}) and (\ref{eq:pressure_NLO})) or in the right
hand side of the equation (eq.~(\ref{eq:fluctuations})) for $\beta$,
are divergent when $x^0\to+\infty$ as illustrated in the figure
\ref{fig:I}.
In this plot, one can check that the envelope of the oscillations
grows exponentially, with a growth rate $\lambda\approx 2*\mu_{\rm
  max}$ where $\mu_{\rm max}$ is the maximal Lyapunov exponent in the
resonance band. If the integral in eq.~(\ref{eq:I}) is evaluated with
an upper cutoff that excludes the resonance band from the integration
domain, then $I(x^0)$ grows only linearly, because now its behavior is
dominated by the soft fluctuation modes whose growth is linear.

Even though secular divergences in integrals such as eq.~(\ref{eq:I})
are present in eq.~(\ref{eq:edens_NLO}), they cancel in the
calculation of $\epsilon_{_{\rm NLO}}$ because the energy density in
our toy model is protected by the conservation of the energy momentum
tensor. However, they do not cancel in $p_{_{\rm NLO}}$ which explains
the divergent behavior displayed in fig.~\ref{fig:NLO}.

\subsection{Alternate form of $T^{\mu\nu}$ at NLO}
The secular divergence of the pressure at NLO suggests that the weak
coupling series for the pressure may be better behaved if one develops
a resummation scheme that captures the physics of the secular terms by
identifying their contribution and summing them to all orders in
perturbation theory. Before we do this, we shall discuss a general
formulation of the energy-momentum tensor at NLO which will help
formulate the problem of resumming secular terms.

In previous works~\cite{GelisV2,GelisV3}, we showed that the problem
of computing NLO corrections for {\it inclusive} quantities-such as
components of the energy momentum tensor in field theories with strong
sources could be formulated as an initial value problem. Specifically,
for the energy-momentum tensor, we can write the NLO contribution at
an arbitrary space-time point as the action of a functional operator
acting on the LO contribution,
\begin{equation}
T^{\mu\nu}_{_{\rm NLO}}(x)
=
\Big[
\int d^3\u\; \beta\cdot{\mathbbm T}_\u
+\frac{1}{2}\int d^3\u d^3\v\;\int\frac{d^3\k}{(2\pi)^3 2k}
[a_{+\k}\cdot{\mathbbm T}_\u][a_{-\k}\cdot{\mathbbm T}_\v]
\Big]
T^{\mu\nu}_{_{\rm LO}}(x)
\; ,
\label{eq:NLO-1}
\end{equation}
The operator ${\mathbbm T}_\u$ that appears in eq.~(\ref{eq:NLO-1}) is
the generator of shifts of the initial conditions
$\varphi_0,\partial_0\varphi_0$ (at $x^0=0$) of the classical field, 
\begin{equation}
a\cdot{\mathbbm T}_\u
\equiv
a(0,\u)\frac{\delta}{\delta\varphi_0(\u)}
+
\dot{a}(0,\u)\frac{\delta}{\delta\partial_0\varphi_0(\u)}\; .
\end{equation}
The factor $T^{\mu\nu}_{_{\rm LO}}$ in the functional formulation of
eq.~(\ref{eq:NLO-1}) should therefore be considered as a functional of
the value of $\varphi,\dot\varphi$ at $x^0=0$. The full content of the
temporal NLO evolution of $T^{\mu\nu}$ is contained in
eq.~(\ref{eq:NLO-1}). One can check that this expression is exactly
equivalent to eq.~(\ref{eq:NLO})~\cite{GelisLV3}.

The expression in eq.~(\ref{eq:NLO-1}) has been obtained by splitting
the time evolution at $x^0=0$ such that the $x^0<0$ part of the time
evolution is described by the operator in the square brackets, and the
evolution at $x^0>0$ is hidden in the functional dependence of
$T^{\mu\nu}_{_{\rm LO}}$ with respect to the value of the classical
field $\varphi$ at $x^0=0$. The choice of $x^0=0$ for this split in
the time evolution is arbitrary and equivalent formulas can be
obtained with other choices\footnote{The splitting of the time
  evolution in two halves need not be done at a constant $x^0$ and any
  locally space-like hypersurface will suffice.}.  Here, our choice is
motivated by the fact that $x^0$ is the time at which the
external source $J$ turns off.  In view of the resummation we will use
later, it is important to note that the quantum field fluctuations
$\beta$ and $a_{\pm\k}$ are still small relative to the classical
field at the splitting time used in the formula. That this is true in
our case is transparent from the figure~\ref{fig:NLO}.

\subsection{Resummation of the NLO corrections}
As seen previously, the fixed order NLO calculation is not meaningful
after a certain time, because it gives a pressure that is larger than
the LO contribution. The NLO contribution (and likely any higher fixed
loop order contribution) has secular divergences because it involves
the {\sl linearized} equation of motion for perturbations to the
classical field $\varphi$. In other words, if $\psi\equiv\varphi+a$,
the NLO calculation approximates the dynamics of $\psi$ by
\begin{eqnarray}
\square\varphi+V^\prime(\varphi)&=&J\nonumber\\
\Big[\square+V^{\prime\prime}(\varphi)\Big]\,a&=&0\; ,
\label{eq:classical+quant}
\end{eqnarray}
on the grounds that the nonlinear terms in $a$ are formally of higher
order in $g^2$. Obviously, if the dynamics of $\psi$ was treated
exactly, by solving instead\footnote{Though this expression looks
  identical to the first equation of eq.~(\ref{eq:classical+quant}),
  the initial conditions for this equation are different, leading to a
  different solution.}
\begin{equation}
\square\psi+V^\prime(\psi)=J\; ,
\end{equation}
we would not have any divergence because the $\psi^4$ potential would
prevent runaway growth of $\psi$. However, in order to achieve this
substitution, we must include in our calculation some contributions
that are of higher order in $g^2$. Thus, we seek a resummation that
restores the lost nonlinearity in the field fluctuations, while
keeping in full the LO and NLO contributions that we have already
calculated.

As we will argue in this section, a simple resummation that leads to
an energy-momentum tensor which is finite at all times consists in
starting from eq.~(\ref{eq:NLO-1}) and in exponentiating the operator
inside the square brackets,
\begin{equation}
T^{\mu\nu}_{\rm resum}(x)
\!\equiv\!
\exp\!\Big[
\int \!\!d^3\u\, \beta\cdot{\mathbbm T}_\u
+\frac{1}{2}\!\int\! d^3\u d^3\v\!\int\!\frac{d^3\k}{(2\pi)^3 2k}
[a_{+\k}\cdot{\mathbbm T}_\u][a_{-\k}\cdot{\mathbbm T}_\v]
\Big]
T^{\mu\nu}_{_{\rm LO}}(x)
\, ,
\label{eq:sum}
\end{equation}
If we Taylor expand the exponential, we recover the full
expressions for the LO and NLO contributions, plus an infinite series
of other terms that are of higher order in $g^2$,
\begin{equation}
T^{\mu\nu}_{\rm resum}(x)
=
\frac{Q^4}{g^2}
\Big[
\underbrace{c_0+c_1 g^2}_{\mbox{fully}} +\underbrace{c_2 g^4+\cdots}_{\mbox{partly}}
\Big]\; .
\end{equation}
From the form of eq.~(\ref{eq:sum}), it is not evident that the
exponentiation leads to a better behaved result; on the surface it appears
that we are including an infinite series of terms that are
increasingly pathological at large times. To see that the result is
now stable when $x^0\to+\infty$, let us consider some generic function
of the classical field at the point $x$, ${\tilde F}[\varphi(x)]$.
The field $\varphi(x)$ is itself a functional of the
values\footnote{Although we do not write that explicitly in order to
  simplify the notation, $\varphi_0$ and $\dot\varphi_0$ may depend on
  the position $\x$.} $\varphi_0$ of the field and $\dot\varphi_0$ of
its first time derivative at $x^0=0$.  Thus, the quantity ${\tilde
  F}[\varphi(x)]$ is implicitly a function of
$\varphi_0,\dot\varphi_0$,
\begin{equation}
{\tilde F}[\varphi(x)] \equiv F[\varphi_0,\dot\varphi_0]\; .
\end{equation}
 Note now that the exponential of $\beta\cdot{\mathbbm
  T}_\u$ is a translation operator when it acts on a functional
$F[\varphi_0(\u),\dot{\varphi}_0(\u)]$,
\begin{equation}
\exp\Big[\int d^3\u\; 
\beta\cdot{\mathbbm T}_\u\Big]\,F[\varphi_0,\dot{\varphi}_0]
=
F[\varphi_0+\beta,\dot{\varphi}_0+\dot{\beta}]\; .
\label{eq:ident-1}
\end{equation}
The first term in the exponential in eq.~(\ref{eq:sum}) therefore merely
shifts the initial conditions $\varphi_0,\dot\varphi_0$ at $x^0=0$ of
the classical field $\varphi$ (by amounts $\beta,\dot\beta$).
Similarly, the second term, that involves the exponential of an
operator that has two ${\mathbbm T}$'s, can be rewritten as a sum over
fluctuations of the initial classical field\footnote{An elementary
  form of the identity,
\begin{equation*}
  e^{\frac{\gamma }{2}\partial_x^2}\,f(x)
  =
  \int_{-\infty}^{+\infty}dz\;
  \frac{e^{-z^2/2\gamma }}{\sqrt{2\pi\gamma }}\,f(x+z)
  \; ,
\end{equation*}
can be proven by doing a Taylor expansion of the exponential in the
left hand side and of $f(x+z)$ in the right hand side. In this simple
example, one sees that an operator which is Gaussian in derivatives is
a {\sl smearing operator} that amounts to convoluting the target
function with a Gaussian. Another way of proving the formula is to
apply a Fourier transform to both sides of the equation.}
\begin{eqnarray}
&&
\exp\Big[
\frac{1}{2}\!\int\! d^3\u d^3\v\int\!\frac{d^3\k}{(2\pi)^3 2k}
[a_{+\k}\cdot{\mathbbm T}_\u][a_{-\k}\cdot{\mathbbm T}_\v]
\Big]\,F[\varphi_0,\dot{\varphi}_0]
=
\nonumber\\
&&\qquad\qquad=
\int [D\alpha D\dot\alpha]\,Z[\alpha,\dot\alpha]\,
F[\varphi_0+\alpha,\dot{\varphi}_0+\dot{\alpha}]\; ,
\label{eq:ident-2}
\end{eqnarray}
where the distribution $Z[\alpha,\dot{\alpha}]$ is Gaussian in
$\alpha(\x)$ and $\dot\alpha(\x)$, with 2-point correlations given by
\begin{eqnarray}
\big<\alpha(\x)\alpha(\y)\big>&=&
\int\frac{d^3\k}{(2\pi)^3 2k}\;a_{+\k}(0,\x)a_{-\k}(0,\y)\; ,
\nonumber\\
\big<\dot\alpha(\x)\dot\alpha(\y)\big>&=&
\int\frac{d^3\k}{(2\pi)^3 2k}\;\dot{a}_{+\k}(0,\x)\dot{a}_{-\k}(0,\y)\; .
\label{eq:gaussian}
\end{eqnarray}
Therefore, the energy-momentum tensor resulting from the resummation of
eq.~(\ref{eq:sum}) can be written as
\begin{eqnarray}
T_{\rm resum}^{\mu\nu}
=
\int [D\alpha(\x) D\dot\alpha(\x)]\,Z[\alpha,\dot\alpha]\;
T^{\mu\nu}_{_{\rm LO}}[\varphi_0+\beta+\alpha]\; ,
\label{eq:sum1}
\end{eqnarray}
where $T^{\mu\nu}_{_{\rm LO}}[\varphi_0+\beta+\alpha]$ denotes the LO
energy-momentum tensor evaluated with a {\sl classical field} whose
initial condition at $x^0=0$ is $\varphi_0+\beta+\alpha$ (and likewise
for the first time derivative).

From eq.~(\ref{eq:sum1}), one can now see why the proposed resummation
cures the pathologies of the NLO contribution. While the fixed-order
NLO result involved linearized perturbations to the classical fields
(that are generically divergent when $x^0\to \infty$), in the resummed
expression these perturbations appear only as a shift of the initial
condition for the full {\sl non-linear equation of motion}. After this
resummation, the evolution of the perturbations at $x^0>0$ is no
longer linear--since the $\phi^4$ potential is bounded from below the
evolution is stable.

In addition to manifestly demonstrating the stable evolution demanded
by the underlying theory, eq.~(\ref{eq:sum1}) is a most useful
expression for a practical implementation of our resummation. It is
important to note however that the integral over $\k$ in the 2-point
correlations (eq.~(\ref{eq:gaussian})) that define the Gaussian
distribution of $\alpha$ and $\dot{\alpha}$ should be cut-off at a
value $\Lambda\sim g\varphi_0\sim Q$ in order to avoid ultraviolet
singularities. With such a cutoff, one can show that the sensitivity
to the value of the cutoff is of higher order in $g^2$, while at the
same time being large enough to include in the resummation all the
relevant unstable modes (the modes with $Q\lesssim |\k|$ are all
stable).

\section{$T_{\rm resum}^{\mu\nu}$ from spatially uniform fluctuations}
\label{sec:toy}
Before we proceed to a full 3+1-dimensional numerical evaluation of
eq.~(\ref{eq:sum1}) with an {\it ab initio} computation of
eq.~(\ref{eq:gaussian}), we shall first consider, as a warm-up
exercise, a computation including only spatially homogeneous
fluctuations. Albeit not realistic, this much simpler calculation will
be very instructive in understanding the effects of these fluctuations
on the behavior of the energy-momentum tensor.

\subsection{Setup of the problem}
For spatially homogeneous fluctuations, the main
simplification is that functional integrations over the fields
$\alpha$ and $\dot\alpha$ in eq.~(\ref{eq:sum1}) become ordinary
integrals over a pair of real numbers, with the Gaussian weight
\begin{equation}
  Z(\alpha,\dot\alpha)\equiv 
  \exp\left[
    -\left(
        \frac{\alpha^2}{2\sigma_1}+\frac{\dot\alpha^2}{2\sigma_2}
      \right)
    \right]\; .
\label{eq:Zuniform}
\end{equation}
The two parameters $\sigma_{1,2}$ can be used in this toy calculation
to control the magnitude of the fluctuations.  In the limit
$\sigma_{1,2}\to 0$, we recover the leading order result which of
course receives no contribution from the fluctuations.

The second important simplification in this toy calculation is that
since both the underlying classical field and the fluctuations are
spatially homogeneous, the field equation of motion is an ordinary
differential equation\footnote{\label{foot:jacobi}In this case, the
  field equations can even be solved analytically. This can be
  seen very simply: from energy conservation,$\frac{1}{2}\,
  {\dot\varphi}^2 + V(\varphi)=E_0= V(\phi_{\rm max})$ (with $\phi_{\rm max}$
  the amplitude of the oscillations of $\varphi(t)$), on gets
\begin{equation*}
t = {\rm const}+
\frac{1}{\sqrt{2}} 
\int_{0}^{\varphi(t)}
\frac{d\psi}{\sqrt{V(\phi_{\rm max}) - V(\psi)}} \; .
\end{equation*}
For a $\phi^4$ potential, the integral in the right hand side is an
elliptic integral, and one can express $\varphi(t)$ as
\begin{equation*}
  \varphi(t) = \phi_{\rm max} 
\;{\rm cn}_{1/2}\,( g\phi_{\rm max} t/\sqrt{24} +{\rm const})
\; ,
\end{equation*}
where ${\rm cn}_{1/2}$ is the Jacobi elliptic function of the first
kind with the elliptic modulus $k=1/2$. This expression is periodic
with a period $T = {2\sqrt{24}} K(1/2)/{g\phi_{\rm max}}$, where
$K(1/2)\approx 1.85$ is the complete elliptic integral of the first
kind.}.  One should note that the characteristic oscillation
frequency is directly proportional to the amplitude of $\varphi(t)$.
This property of the solution will be key in interpreting the results
that follow.

We now turn to the computation of the resummed pressure and energy
density in this toy model.  From eqs.~(\ref{eq:edens-LO}) and
(\ref{eq:sum1}), the expressions for the energy density and the
pressure read
\begin{eqnarray}
  \epsilon_{\rm resum}&=&\left<\frac{1}{2}\dot\varphi^2+V(\varphi)\right>_{\alpha,\dot\alpha}\; ,\nonumber\\
  p_{\rm resum}&=&\left<\frac{1}{2}\dot\varphi^2-V(\varphi)\right>_{\alpha,\dot\alpha}\; ,
\end{eqnarray}
where $\varphi$ is the solution of the classical equation of motion
whose value at $x^0=0$ is $\varphi_0+\alpha$ and whose time derivative
at $x^0=0$ is $\dot\varphi_0+\dot\alpha$. The brackets
$\big<\cdots\big>_{\alpha,\dot\alpha}$ denote an averaging over all
possible values of $\alpha,\dot\alpha$ with the distribution of
eq.~(\ref{eq:Zuniform}).

\subsection{Energy momentum tensor}
\begin{figure}[htbp]
\begin{center}
\resizebox*{8cm}{!}{\rotatebox{-90}{\includegraphics{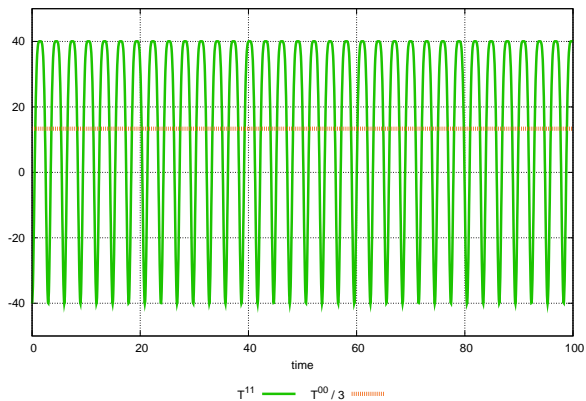}}}
\end{center}
\caption{\label{fig:Tmunu-LO} Components of $T_{\rm LO}^{\mu\nu}$,
  where when no quantum fluctuations are included.}
\end{figure}

\begin{figure}[htbp]
\begin{center}
\resizebox*{8cm}{!}{\rotatebox{-90}{\includegraphics{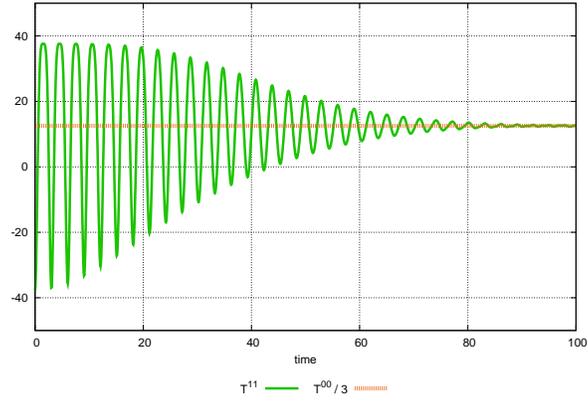}}}
\end{center}
\caption{\label{fig:uniform}Components of $T_{\rm resum}^{\mu\nu}$ obtained with a
  Gaussian ensemble of spatially uniform quantum fluctuations.}
\end{figure}

In fig.~\ref{fig:Tmunu-LO}, we display the result of the toy model
calculation in the limit where we do not have fluctuations
($\sigma_{1,2}\to 0$). As anticipated, the result is equivalent to the
one displayed in fig.~\ref{fig:LO} for the leading order
calculation.
In this figure, for reasons that will become obvious shortly, we have
represented the energy density divided by three.  In
fig.~\ref{fig:uniform}, we show the results of the same calculation
performed with non-zero widths $\sigma_{1,2}$ for the Gaussian
distribution of fluctuations.
We observe a striking difference of the resummed result compared to
the previous (LO) figure--the oscillations of the pressure are damped
and the value of the pressure relaxes to $\epsilon/3$. Subsequently,
one has a single-valued relationship between the pressure and the
energy density, namely, an {\sl equation of state} -- specifically, the
equation of state $\epsilon=3p$ of a scale invariant system in $1+3$
dimensions.

\subsection{Phase-space density}
\begin{figure}[htbp]
\begin{center}
\resizebox*{8cm}{!}{\rotatebox{-90}{\includegraphics{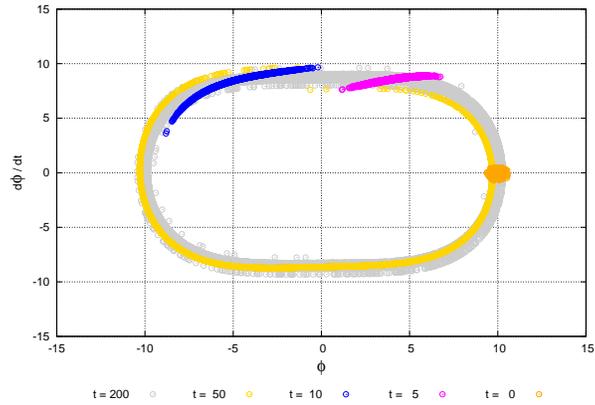}}}
\end{center}
\caption{\label{fig:ps}Phase-space distribution of the ensemble of
  classical fields at various stages of the time evolution.}
\end{figure}
It is also instructive to look at the phase-space density
$\rho_t(\varphi,\dot\varphi)$ of the points $(\varphi,\dot\varphi)$ as the
system evolves in time. This is shown in fig.~\ref{fig:ps}. At
$t=0$, we start with a Gaussian distribution of the initial
conditions, with a small dispersion around the average values
($\varphi=10$ and $\dot\varphi=0$ in our example).

Each initial condition then evolves independently according to the
classical equation of motion, and the corresponding trajectory in the
$(\varphi,\dot\varphi)$ plane is a closed loop\footnote{These loops
  are constant energy curves $\frac{1}{2}\dot\varphi^2+V(\varphi)=H$.}
due to the periodicity of classical solutions. One observes that the
initially Gaussian-shaped cloud of points starts spreading around a
closed loop, to eventually fill it entirely when $x^0\to+\infty$. When
this asymptotic regime is reached, the density
$\rho_t(\varphi,\dot\varphi)$ depends only on the energy--roughly
speaking, the radial coordinate in the plot of figure~\ref{fig:ps}
and no longer on the angular coordinate.

A more formal way of phrasing the same result is to first note that
the time evolution of the phase-space density $\rho_t$ obeys the
Liouville equation,
\begin{equation}
\frac{\partial\rho_t}{\partial t}+\{\rho_t,H\} =0\; ,
\end{equation}
where $\{\cdot,\cdot\}$ is the classical Poisson bracket. Therefore,
if a stationary distribution is reached at late times, it can only
depend on $\varphi$ and $\dot\varphi$ via $H(\varphi,\dot\varphi)$.
The asymptotic behavior of the phase-space density in our toy model is
reminiscent of a micro-canonical equilibrium state, in which the
phase-space density is uniform on a constant energy
manifold\footnote{It should be noted here that a spatially homogeneous
  field is very special regarding this issue; indeed, any non-linear
  system with a single degree of freedom is ergodic. This is not
  necessarily the case if there are more than one degrees of freedom,
  as is the case in a full fledged field theory.}. In other words, all
micro-states that have the same energy are equally likely.

\subsection{Interpretation of the results}
We shall now discuss the physical interpretation of our results, first
discussing the decoherence of the temporal evolution of the fields and
their time derivatives, and subsequently, the impact of decoherence on
the relaxation of the pressure towards that of a scale invariant system.
\subsubsection{Decoherence time}
Of the previous numerical observations, the easiest to understand is
the spreading of the phase-space density around a closed orbit.
Because the oscillations are non-harmonic, the various points in the
plot of figure \ref{fig:ps} rotate at different speeds\footnote{The
  assumption of a scale invariant theory simplifies some expressions
  here, but is not crucial to the argument. The only requirement for
  this phenomenon is that the frequency of the oscillations depends on
  their amplitude; thus any non-harmonic potential will lead to
  similar results.}; in a $\phi^4$ potential, the outer points rotate
faster than the inner ones. Therefore, as time increases, the cloud of
points spreads more and more due to this effect.

One can estimate the time necessary for the cloud of points to spread
over a complete orbit. This happens when the angular spread of the
points reaches the value $2\pi$. For one field configuration, this
angular variable is, up to a phase that depends on the initial
condition, $\theta = \omega t$, and the angular velocity $\omega$
depends only on the energy of that particular field configuration. (In our case, this phase is small for a narrow Gaussian distribution.) If
we consider two field configurations, their angular variable
difference $\Delta\theta$ increases linearly in time, $\Delta\theta
= \Delta\omega\, t$, where $\Delta\omega$ is the difference between
their angular velocities. In the case of a $g^2\phi^4/4!$ potential,
one can prove that (see the footnote \ref{foot:jacobi})
\begin{equation}
\omega = 
\frac{\pi}{2\sqrt{3}} 
\frac{g\phi_{\rm max}}{\int_{-1}^{+1}\frac{dx}{\sqrt{1-x^4}}}
\approx 0.346\,g\phi_{\rm max}\; ,
\end{equation}
where $\phi_{\rm max}$ is amplitude of the oscillations of the $\varphi$ field.
Thus, the angular shift between the two field configurations is also
$\Delta\theta \approx 0.346\,g\Delta\phi_{\rm max}\, t$, and this
shift reaches $2\pi$ in a time
\begin{equation}
t\approx \frac{18.2}{g\Delta\phi_{\rm max}}\; .
\label{eq:t-decoherence}
\end{equation}
After this time, the two fields have become completely incoherent. We
see that this time is inversely proportional to the coupling constant
$g$, and to the difference of the field amplitudes. Thus a narrow
initial Gaussian distribution will need a longer time to spread around
the orbit than a broader initial distribution.

\subsubsection{Equation of state from quantum averaging}
\label{sec:qavg}
Once we know that the phase-space density spreads uniformly on
constant energy curves, it is easy to understand why the pressure
relaxes towards $\epsilon/3$ when we let the initial conditions for
the classical field fluctuate. The trace of the energy-momentum tensor
(assuming 4 dimensions of space-time) is
\begin{equation}
T^\mu{}_\mu=
\varphi\left(\square\varphi+4\frac{V(\varphi)}{\varphi}\right)
-
\partial_\alpha(\varphi\partial^\alpha\varphi)\; .
\end{equation}
A scale invariant theory in four dimensions is a theory in which the
interaction potential obeys $V^\prime(\varphi)=4V(\varphi)/\varphi$. This is
the case of a $\phi^4$ interaction. Therefore, the first term in the
right hand side of the previous equation vanishes thanks to the
equation of motion of the classical field $\varphi$. This result shows
that the energy-momentum tensor of a single configuration of classical
field is not zero in our model, but is a total
derivative\footnote{There is an alternative ``improved'' definition
  of the energy-momentum tensor that is explicitly
  traceless~\cite{CallaCJ1}.  However, while the energy density has a
  single valued relation to the pressure, this pressure is not the
  canonical pressure.  As we shall discuss later in section
  \ref{sec:ccj}, both definitions give a deviation from ideal
  hydrodynamic flow, which is cured by the quantum averaging described
  here.}. In our simplified toy model where the fields are spatially
homogeneous, the previous relation simplifies to
\begin{equation}
T^\mu{}_\mu=
-
\frac{d(\varphi\dot\varphi)}{dt}\; .
\end{equation}
When averaged over one period, the trace of the energy-momentum of
one classical field configuration vanishes because the classical
field is a periodic function of time,
\begin{equation}
\overline{T^\mu{}_\mu}\equiv \frac{1}{T}\int_t^{t+T}d\tau\;
T^\mu{}_\mu(\varphi(\tau),\dot\varphi(\tau))=0\; ,
\label{eq:time-avg}
\end{equation}
where the result is independent of $t$.  When we calculate the
energy-momentum tensor averaged over fluctuations of the initial
conditions, we are in fact performing an ensemble average weighted by
the phase-space density $\rho_t(\varphi,\dot\varphi)$,
\begin{equation}
\left<T^\mu{}_\mu\right>_{\alpha,\dot\alpha}
=
\int d\varphi\, d\dot\varphi\; \rho_t(\varphi,\dot\varphi)\;T^\mu{}_\mu(\varphi,\dot\varphi)\; ,
\end{equation}
and the time dependence of the left hand side comes from that of the
density $\rho_t$. 
It is convenient to
trade the integration variables $\varphi,\dot\varphi$ for energy/angle
variables $E,\theta$,
\begin{equation}
\left<T^\mu{}_\mu\right>_{\alpha,\dot\alpha}
=
\int dE d\theta\; \tilde\rho_t(E,\theta)\;
T^\mu{}_\mu(E,\theta)\; ,
\end{equation}
where $\tilde\rho_t$ is the phase-space density in the new system of
coordinates\footnote{$\tilde\rho_t$ is equal to the original $\rho_t$ times the
  Jacobian of the change of variables.}.  Our first result shows that
$\tilde\rho_t(E,\theta)\longrightarrow \tilde\rho_t(E)$, namely,
becomes independent of $\theta$ at late times, which enables us to
write the previous equation as
\begin{equation}
  \left<T^\mu{}_\mu\right>_{\alpha,\dot\alpha}
  \empile{\approx}\over{t\to \infty}
  \int dE \; \tilde\rho_t(E)\;\int d\theta\;
  T^\mu{}_\mu(E,\theta)\; .
\end{equation}
The crucial point here is that the integral over $\theta$ is simply
the integral over one orbit for a single classical field configuration
(eq.~(\ref{eq:time-avg})),
\begin{equation}
\int d\theta\;
  T^\mu{}_\mu(E,\theta)
=
\frac{2\pi}{T}\int_t^{t+T}d\tau\;
T^\mu{}_\mu(\varphi(\tau),\dot\varphi(\tau))=0\; .
\end{equation}
Thus, we have proven that
\begin{equation}
  \epsilon-3p=\left<T^\mu{}_\mu\right>_{\alpha,\dot\alpha}
  \empile{\approx}\over{t\to \infty}0\; ,
\end{equation}
in agreement with what we have observed numerically.  Moreover, from
the derivation of this result, it is clear that the time necessary to
reach this limit is the same as the time (in
eq.~(\ref{eq:t-decoherence})) necessary for the phase-space density to
become independent of the angular variable $\theta$.

\subsection{Effect of the longitudinal expansion}

We have thus far considered a system of strong fields enclosed in a
box of fixed volume. There is therefore no concept of hydrodynamical
flow in such a system. To fully understand the implications of
decoherence and relaxation of the pressure we have discussed
previously for hydrodynamical flow, we will now simply generalize
the toy problem of spatially uniform fields and fluctuations to a
system undergoing a boost invariant one dimensional expansion.

\subsubsection{Relaxation of the pressure}

The geometry of the one dimensional expansion (chosen to be the $z$
direction) is appropriate to describe the collision of two projectiles
(nuclei) at ultrarelativistic energies.  The natural coordinates are
the proper time $\tau$ and rapidity $\eta$ defined by
\begin{eqnarray}
  \tau&\equiv& \sqrt{t^2-z^2}\; ,\nonumber\\
  \eta&\equiv& \frac{1}{2}\ln\left(\frac{t+z}{t-z}\right)\; .
\end{eqnarray}
In the spatial plane orthogonal to the $z$ axis, the coordinates are
denoted by $\x_\perp$.  In this system of coordinates, the classical
equation of motion for a field $\varphi$ that depends only on proper
time is
\begin{equation}
\ddot\varphi+\frac{1}{\tau}\dot\varphi+\frac{g^2}{6}\varphi^3=0\;,
\label{eq:eom-expanding}
\end{equation}
where the dot now denotes a derivative with respect to $\tau$.  The
analog of the the toy problem we discussed previously in the first part of
this section is to let the initial conditions of the field have
Gaussian fluctuations $\alpha,\dot\alpha$ that are independent of
$\eta$ and $\x_\perp$.

\begin{figure}[htbp]
\begin{center}
\resizebox*{8cm}{!}{\rotatebox{-90}{\includegraphics{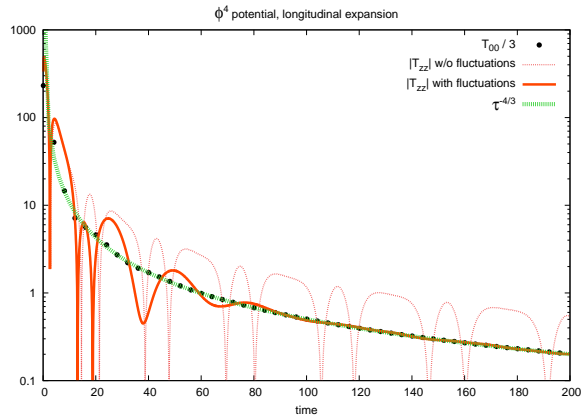}}}
\end{center}
\caption{\label{fig:exp-phi4}Numerical evaluation of $T^{\mu\nu}$ for
  fields undergoing a boost invariant 1-dimensional expansion in a
  $\phi^4$ theory with a Gaussian ensemble of spatially uniform
  initial fluctuations.}
\end{figure}

The components of the energy-momentum tensor in this system of
coordinates, averaged over the fluctuations of the initial conditions are
\begin{eqnarray}
\epsilon\equiv T^{\tau\tau}
&=&
\Big<\frac{1}{2}\dot\varphi^2+V(\varphi)\Big>_{\alpha,\dot\alpha}\; ,
\nonumber\\
p\equiv T^{xx}=T^{yy} = \tau^2 T^{\eta\eta}
&=&
\Big<\frac{1}{2}\dot\varphi^2-V(\varphi)\Big>_{\alpha,\dot\alpha}\; .
\end{eqnarray}
As in the fixed volume case, we shall use the distribution of
eq.~(\ref{eq:Zuniform}) for $\alpha,\dot\alpha$.  The result of this
computation is shown in the figure \ref{fig:exp-phi4}.
The dots represent the energy density divided by 3, and one observes
that its time dependence is well described by a $\tau^{-4/3}$ decay
characteristic of boost invariant flow in ideal relativistic
hydrodynamics. If we do not include fluctuations of the initial
conditions, we observe that the pressure oscillates between positive
and negative values, with a decreasing envelope. Conversely, if we
average over an ensemble of initial conditions, we see the
oscillations of the pressure dampen quickly, ensuring that the
pressure approaches one third of the energy density.

\begin{figure}[htbp]
\begin{center}
\resizebox*{8cm}{!}{\rotatebox{-90}{\includegraphics{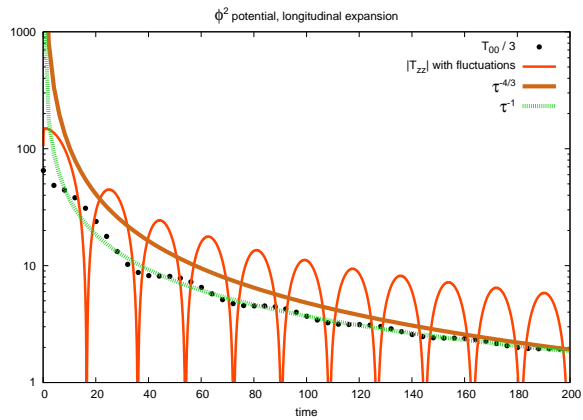}}}
\end{center}
\caption{\label{fig:exp-phi2}Numerical evaluation of $T^{\mu\nu}$ for
  fields undergoing a boost invariant 1-dimensional expansion in a
  $\phi^2$ theory, with a Gaussian ensemble of uniform initial
  fluctuations.}
\end{figure}

These results are in sharp contrast to what one obtains for a $\phi^2$
potential. The results are shown in fig.~\ref{fig:exp-phi2}. In this
case, the fluctuations do not make the pressure converge to
$T^{00}/3$, and the latter decreases as $\tau^{-1}$ instead of
$\tau^{-4/3}$.

\subsubsection{Interpretation of the results for expanding fields}
From the equation of motion in eq.~(\ref{eq:eom-expanding}), we obtain 
\begin{eqnarray}
\frac{d\epsilon}{d\tau} &=&
\dot\varphi\Big[\ddot\varphi+V^\prime(\varphi)\Big] 
= -\frac{1}{\tau}\dot\varphi^2\nonumber\\
&=& -\frac{\epsilon+p}{\tau}\; .
\label{eq:exp-hydro}
\end{eqnarray}
This equation, which is valid for individual classical field
configurations at every time, is identical to Euler's equation for
boost-invariant ideal hydrodynamics.  The difference with
hydrodynamics lies in the fact that hydrodynamics assumes the
existence of an equation of state $p=f(\epsilon)$ to ensure a closed
form expression in eq~(\ref{eq:exp-hydro}). In classical field
dynamics, one is not free to impose a relationship between $\epsilon$
and $p$ since they are both completely determined from the field
$\varphi$ and its derivative $\dot\varphi$. For instance, as seen in
fig.~\ref{fig:exp-phi4}, for a single classical solution, we do not
have a one-to-one correspondence between $\epsilon$ and $p$;
$\epsilon$ has a monotonous behavior while $p$ oscillates. What is
remarkable is that the ensemble average over the initial conditions
leads in a short time to a one-to-one correspondence $\epsilon=3p$,
which is precisely the equation of state one would use in boost
invariant hydrodynamics of a perfect fluid.

The mechanism whereby this relationship is reached is the same as in
the non-expanding case. As previously, one can prove that for a single
phase-space trajectory, the time averages of $\epsilon$ and $p$ obey a
relation identical to the expected equation of state, because the
trace of the energy momentum tensor is a total derivative.  Then, by
using the fact that different initial conditions lead to different
oscillation frequencies, one gets the phase decoherence that enables
us to transform the ensemble average over the initial conditions into
a time average along one classical field trajectory. This decoherence
is the missing ingredient in the harmonic case-as we noted previously,
it arises for the $\phi^4$ theory because the angular velocity of the
phase space trajectory of an individual configuration depends on the
amplitude of the configuration.

From this result, it is very easy to obtain the $\tau^{-4/3}$ behavior of
the energy density. The ensemble
average of eq.~(\ref{eq:exp-hydro}) at late times is 
\begin{equation}
\frac{d\epsilon}{d\tau}\empile{=}\over{\tau\to+\infty}
-\frac{4}{3}\epsilon\; ,
\end{equation}
which leads immediately to the observed behavior. Since both
$\epsilon$ and $p$ decrease like $\tau^{-4/3}$ even for a single
configuration (if one considers the envelope of the oscillations of
$p$), this means that at late times we have
\begin{equation}
\varphi\sim \tau^{-1/3}\; ,\quad \dot\varphi \sim \tau^{-2/3}\; .
\end{equation} 
This behavior is seen from the simple ansatz $\varphi(\tau)\sim
\cos(f(\tau)) \tau^{-1/3}$, which, while inaccurate in detail,
qualitatively captures the right physics. For a $\varphi^4$ potential,
the frequency $\dot{f}\propto \varphi\sim \tau^{-1/3}$ for a $\phi^4$
potential; using this relation in our ansatz gives the stated result.
The fact that $\dot\varphi$ decreases faster than $\varphi$ is due to
the slowing down of the oscillations with time, as their amplitude
decreases.

\subsubsection{Callan-Coleman-Jackiw energy-momentum tensor}
\label{sec:ccj}
An alternate definition of the energy-momentum tensor was proposed by
Callan, Coleman and Jackiw (CCJ)~\cite{CallaCJ1}. Their expression is
explicitly traceless. They argued further that their form of the
stress energy tensor improved properties relative to the canonical one
with regard to renormalization. Let us briefly summarize the
differences between the usual definition of $T^{\mu\nu}$ and CCJ's.
With the canonical definition that we have used thus far, one has,
\begin{eqnarray}
T^{\mu\nu} &\equiv& (\partial^\mu\varphi)(\partial^\nu\varphi)-g^{\mu\nu}{\cal L}\nonumber\\
\epsilon&=&\frac{1}{2}\dot\varphi^2+V(\varphi)\nonumber\\
p&=&\frac{1}{2}\dot\varphi^2-V(\varphi)\nonumber\\
\epsilon-3p &=& \frac{d(\varphi\dot\varphi)}{d\tau}\nonumber\\ 
\frac{d\epsilon}{d\tau}&=&-\frac{\epsilon+p}{\tau}\; .
\end{eqnarray}
With this definition of $T^{\mu\nu}$, one obtains the equation for
Bjorken hydrodynamics automatically for each configuration of the
classical field. However, one gets $\epsilon=3p$ only through
decoherence, by averaging over an ensemble of initial conditions.

In comparison, with CCJ's definition of the energy-momentum tensor, one has
\begin{eqnarray}
T^{\mu\nu} &\equiv& (\partial^\mu\varphi)(\partial^\nu\varphi)-g^{\mu\nu}{\cal L}-\frac{1}{6}(\partial^\mu\partial^\nu-g^{\mu\nu}\square)\varphi^2\nonumber\\
\epsilon&=&\frac{1}{2}\dot\varphi^2+V(\varphi)\nonumber\\
p&=&\frac{1}{2}\dot\varphi^2-V(\varphi)-\frac{1}{6}\square\varphi^2\nonumber\\
\epsilon-3p &=& 0\nonumber\\ 
\frac{d\epsilon}{d\tau}&=&-\frac{\epsilon+p}{\tau}-\frac{1}{3\tau}\frac{d(\varphi\dot\varphi)}{d\tau}\; .
\end{eqnarray}
With this form of the energy momentum tensor, the equation of state is
satisfied for each classical field configuration, but not Bjorken's
hydrodynamic equation. It is only after an average over an ensemble of
initial conditions that the last term ($\sim
d(\varphi\dot\varphi)/d\tau$) in the last equation vanishes by
decoherence.

Since one requires simultaneously
\begin{eqnarray}
\frac{d\epsilon}{d\tau}+\frac{\epsilon+p}{\tau}&=&0\nonumber\\
\epsilon-3p&=&0 \; ,
\end{eqnarray}
for ideal hydrodynamical flow, one sees that there is no discrepancy
between the two descriptions despite the apparent differences.  For
our choice of the energy momentum tensor, the reason why we didn't
have an ideal hydrodynamic behavior at the beginning of the evolution
was because of the lack of an equation of state. In the case of CCJ's
energy momentum tensor, it is because of a violation of the canonical
Euler equation. The net effect of the quantum averaging in each case
is to get rid of one or the other violation thereby ensuring ideal
hydrodynamical behavior. In the following section, we will discuss
only the canonical energy momentum tensor, for which the focus is on
obtaining the equation of state as the necessary condition for
hydrodynamical flow.

\section{Results from the full fluctuation spectrum}
In the previous section, we showed that averaging over an ensemble of
initial conditions for classical fields can lead the pressure to relax
towards one third of the energy density. However, this study was
oversimplified since we used only fluctuations that are uniform in
space, and their Gaussian distribution was set by hand. However,
quantum field theory {\sl predicts} what the spectrum of these
fluctuations is: one should average the LO energy-momentum
tensor\footnote{Since, at $x^0=0$, $\beta$ (see eq.~(\ref{eq:sum1})) is
  a small shift that does not fluctuate, we have absorbed it into a
  redefinition of the classical field $\varphi_0$.},
\begin{eqnarray}
T_{\rm resum}^{\mu\nu}
=
\left<
T^{\mu\nu}_{_{\rm LO}}[\varphi_0+\alpha]\right>_{\alpha,\dot\alpha}\; ,
\label{eq:sum1-1}
\end{eqnarray}
over space-dependent random Gaussian fields $\alpha$ and
$\dot\alpha$ that have the following variance:
\begin{eqnarray}
\big<\alpha(\x)\alpha(\y)\big>&=&
\int\frac{d^3\k}{(2\pi)^3 2k}\;a_{+\k}(0,\x)a_{-\k}(0,\y)
\nonumber\\
\big<\dot\alpha(\x)\dot\alpha(\y)\big>&=&
\int\frac{d^3\k}{(2\pi)^3 2k}\;\dot{a}_{+\k}(0,\x)\dot{a}_{-\k}(0,\y)\; ,
\label{eq:gaussian-1}
\end{eqnarray}
which leaves no freedom to handpick what fluctuations we use.  From
these formulas, we can numerically compute {\it ab initio} the
behavior of the pressure. The only tunable quantities in the
calculation are then the scale $Q$ (or more generally the source $J$)
that controls the amount of energy injected into the system at $t<0$,
and the coupling constant $g$. Note that the above spectrum of
fluctuations for the initial condition for the field $\varphi$ is
equivalent to parameterizing the initial field
as\footnote{\label{foot:berry}If we recall that the $a_{\pm\k}$'s are
  plane waves modified by the presence of the background field
  $\varphi_0$, we observe that the fluctuating part of the initial
  field is very similar to the form of the wavefunction of high lying
  eigenstates for quantum systems that obey Berry's conjecture.}
\begin{equation}
\varphi(0,\x)\equiv \varphi_0(0,\x)+\int\frac{d^3\k}{(2\pi)^3 2k}\;
\Big[c_\k\, a_{+\k}(0,\x)+c_\k^*\, a_{-\k}(0,\x)
\Big]\; ,
\label{eq:berry-1}
\end{equation}
where the $c_\k$ are random Gaussian numbers with the following
variance
\begin{equation}
\big<c_\k c_\l\big>=0\; ,\qquad 
\big<c_\k c_\l^*\big>= (2\pi)^3 |\k|\delta(\k-\l)\; .
\label{eq:berry-2}
\end{equation}
Details of the numerical lattice computation are relegated to the
appendix \ref{app:lattice}; in this section we focus on the results
from numerical simulations with this {\it ab initio} spectrum of
fluctuations.

\subsection{Numerical results}
Unless stated otherwise, the numerical results in this section are
obtained on a $12^3$ lattice\footnote{In some instances, we have also
  performed simulations on a $20^3$ lattice and found only very small
  differences as long as the physical scales are below the lattice
  cutoff.}. The functional integration in eq.~(\ref{eq:sum1}) is
approximated by a Monte-Carlo average over 1000 configurations of the
initial conditions, distributed according to eqs.~(\ref{eq:berry-1})
and (\ref{eq:berry-2}).

\begin{figure}[htbp]
\begin{center}
\resizebox*{8cm}{!}{\rotatebox{-90}{\includegraphics{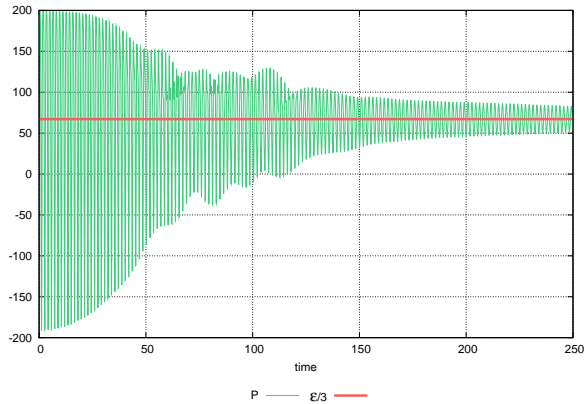}}}
\end{center}
\caption{\label{fig:pressure1}Time evolution of the pressure averaged
  over the initial fluctuations. All the resonant modes are included
  in the simulation. The coupling constant is $g=0.5$.}
\end{figure}

In fig.~\ref{fig:pressure1}, we show the result of the
computation of the pressure averaged over the Gaussian ensemble of
initial conditions, for a value of the coupling\footnote{Since the
  prefactor in the interaction potential is $g^2/4!$, a value $g=0.5$
  corresponds to a very weak coupling strength.} $g=0.5$. We also show the
energy density divided by three on the same plot.
All the quantities in this plot are expressed in lattice units, which
means that the horizontal axis is $t/a$ (where $a$ is the lattice
spacing) and the vertical axis should be understood as $\epsilon
a^4/3$ or $pa^4$. The lattice cutoff in this simulation is chosen to
be just above the upper limit of the parametric resonance window
($k/m_0=3^{-1/4}$ where $m_0^2=g^2\varphi_0^2/2$); therefore, all the
resonant modes take part in the dynamics of the system.

We observe that the ensemble averaged pressure relaxes towards
$\epsilon/3$. This plot, obtained with the spectrum of
fluctuations predicted by quantum field theory, is one of the central
results of this paper. One can qualitatively identify two stages in
this relaxation: (1) in the range $0\le t\lesssim 50$, the amplitude
of the pressure oscillations decreases very quickly to a moderate
value and, (2) from time $50$ onwards, one has a slower approach of
the pressure to $\epsilon/3$ that gets slowly rid of the residual
oscillations. We will observe again this two-stage time evolution when
we look at the fluctuations of the energy density.

\subsection{Influence of the resonant modes}
\begin{figure}[htbp]
\begin{center}
\resizebox*{8cm}{!}{\rotatebox{-90}{\includegraphics{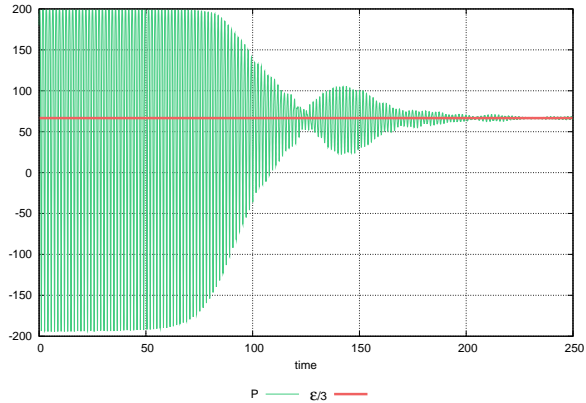}}}
\end{center}
\caption{\label{fig:pressure2}Time evolution of the pressure averaged
  over the initial fluctuations. The lattice cutoff is located below
  the resonance band in order to exclude them from the simulation. The
  coupling constant is $g=0.5$.}
\end{figure}
In section \ref{sec:toy}, we observed that the pressure relaxes
to $\epsilon/3$ even if only the mode $\k=0$ is included in the
simulation. This was understood as an consequence of the phase decoherence
that exists in a non-harmonic potential between classical solutions
that have slightly different amplitudes. When we include all the
$\k$-modes of the fluctuations, the situation becomes more
complicated. In particular, the stability analysis of these
fluctuations (see the appendix \ref{app:stability}) indicates that in
addition to a linear instability of the soft modes due to the above
mentioned decoherence phenomenon, there are also exponentially
unstable modes in a narrow band of values $\k$.

In order to assess the role played in the time evolution by the modes
of the resonance band, we performed a second simulation with the same
physical parameters, but now with the lattice cutoff placed just below
the lower end of the resonance band. This makes certain that none of
the modes that exist on this lattice has an exponential instability.
Since the resonance band is quite narrow, this is a small change of
the cutoff in physical units because the cutoff in the earlier
simulation was just above the upper end of the resonance band.
However, one can see in the figure \ref{fig:pressure2} that excluding
the resonant modes leads to significant changes.

The final outcome,  the relaxation of the pressure towards $\epsilon/3$, is not
changed, but the details of the time evolution of the pressure are
modified. Firstly, one observes a rather long delay during which the
oscillations of the pressure remain almost constant in amplitude.
Then, at a time of order $75$ in lattice units, these oscillations are
damped very quickly to very small wiggles around $\epsilon/3$. Except
for a brief relapse, the oscillations remain very small after this
time. In particular, the two-stage evolution that we observed with the
full spectrum is now replaced by the following two stages: (1) nothing
happens and, (2) very rapid relaxation that leaves almost no residual
oscillations.

Therefore, it appears that the resonant modes, even if their presence
or absence in the resummation does not change the final outcome, do
alter significantly the detailed time evolution of the pressure. At
this point, the precise role of the resonant modes is somewhat
unclear.  It appears that the dynamics of the complete system is much
richer than what one can learn by studying the linearized evolution of
a single mode as done in the stability analysis of the appendix
\ref{app:stability}. This analysis does not capture the non-linear
couplings between the various modes (once the instabilities have made
them large) which gives them a big role in the late stage evolution of
the system. This certainly deserves further study.

\subsection{Dependence on the coupling constant}
The simulation that led to the result of fig.~\ref{fig:pressure1}
was performed with a value $g=0.5$ for the coupling constant -- a very
small value for our scalar field theory since there is also a factor
$1/4!$ in the interaction potential.
\begin{figure}[htbp]
\begin{center}
\resizebox*{8cm}{!}{\rotatebox{-90}{\includegraphics{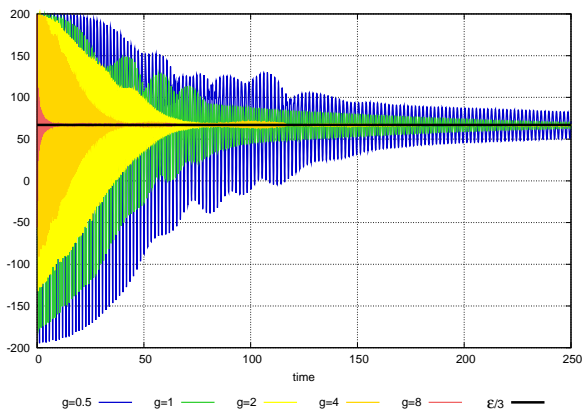}}}
\end{center}
\caption{\label{fig:gdep}Time evolution of the pressure averaged over
  the initial fluctuations for various values of the coupling
  constant: $g=0.5, 1, 2, 4, 8$. All the resonant modes are included
  in the simulation. See footnote ~\ref{footnote:latticeunits}. }
\end{figure}
We have studied the time evolution of the pressure for larger values
of the coupling constant: $g=1,2,4,8$, and the results are shown in
the figure \ref{fig:gdep}. Note that this computation is done at fixed
energy density. Indeed, because $Q$ is the only dimensionful parameter
of our model and there is a factor $1/g$ in the source $J$, the energy
density behaves at leading order as $\epsilon\propto Q^4/g^2$.  Thus,
if we increase $g$ at constant $Q$, the energy density decreases. As
our goal is to assess the time at which the pressure obeys an equation
of state (thereby justifying a hydrodynamical description of the
system), the comparison of the relaxation for various couplings should
be done for systems that have the same energy density. Therefore, in
the comparison shown in fig.~\ref{fig:gdep}, the value of $Q$ has
been adjusted in each simulation such that the energy density remains
unchanged.

Fig.~\ref{fig:gdep} demonstrates that the relaxation time
decreases with increasing coupling constant $g$. 
\begin{figure}[htbp]
\begin{center}
\resizebox*{8cm}{!}{\rotatebox{-90}{\includegraphics{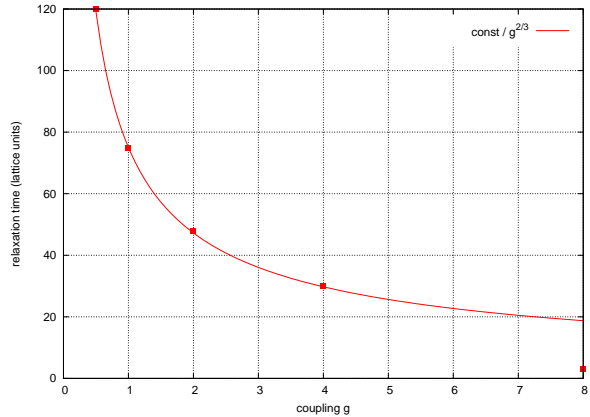}}}
\end{center}
\caption{\label{fig:gdep1}Points: relaxation time (see text for
  the definition used here) as a function of the coupling $g$. Line:
  fit by a power law.}
\end{figure}
In fig.~\ref{fig:gdep1} we have represented the relaxation time, defined
here as the time necessary to reduce the initial oscillations of the
pressure by a factor 4, as a function of the coupling constant $g$ for
our set of values of $g$. One can fit all the points except the last
one ($g=8$) by a power law that suggests the following
dependence\footnote{\label{footnote:latticeunits}The axis of the figure \ref{fig:gdep} are in
  lattice units. Thus, the horizontal axis is $t/a$ and the
  vertical axis $pa^4$ or $\epsilon a^4/3$, where $a$ is the lattice
  spacing. Since our model is scale invariant, the relaxation time
  scales like $\epsilon^{-1/4}$. By eliminating $a$ between the
  horizontal and vertical axis, it is easy to get the value of
  $\epsilon^{1/4}t$. For $g=4$ we have $\epsilon a^4=200$ and the
  relaxation time is $t/a\approx 30$, leading to $\epsilon^{1/4} t
  \approx 113$ (this combination is $11$ for $g=8$). Then, from one's
  favorite value of $\epsilon$ in GeV/fm${}^3$, it is easy to obtain
  the relaxation time in fm's.}
\begin{equation}
t_{\rm relax}=\frac{\rm const}{g^{2/3}\epsilon^{1/4}}\; .
\end{equation}
The right most point in this plot is an outlier that does not follow
this power law, possibly because this value of the coupling is too
extreme for our approximations/resummations to make sense (for $g=8$,
the interaction strength $g^2/4!$ is significantly above 1).

\subsection{Energy density fluctuations}
The results we have shown thus far indicate that the pressure in the
system relaxes towards the equation of state $p=\epsilon/3$, at
relaxation times that decrease as the coupling constant increases.
However, this study does not in and of itself tell us much about the
nature of the state reached by the system. In particular, it does not
tell us whether the system reaches a state of local thermal
equilibrium.
Because we have a system of strong fields whose modes have large
occupation numbers, it is unlikely that the system can be described in
terms of quasi-particles that have a Bose-Einstein distribution. In
section \ref{sec:toy}, we observed that in the simple example studied
that the phase-space density reaches a stationary form reminiscent of
a micro-canonical equilibrium ensemble.  Unfortunately, now that we
are looking at a full fledged quantum field theory, the phase-space is
infinite dimensional and whether the same behavior occurs is difficult
to assess numerically.

There are however signs of thermalization in the fluctuations of the
energy distribution in the system. For the system as a whole, energy
is conserved and will not fluctuate, regardless of whether the system
is in thermal equilibrium or not. However, as is well known for
canonical ensembles, by looking at energy fluctuations in a small
subsystem, one can learn something about the energy exchanges between
this subsystem and the rest of the system which acts as a heat bath.
In particular, the nature of the fluctuations in the energy
distribution of the subsystem can tell us whether it is in equilibrium
with the rest of the system. If this is the case, the fluctuations are
those of a canonical ensemble with a density operator
$\rho\equiv\exp(-\beta H)$.

\begin{figure}[htbp]
\begin{center}
\resizebox*{8cm}{!}{\rotatebox{-90}{\includegraphics{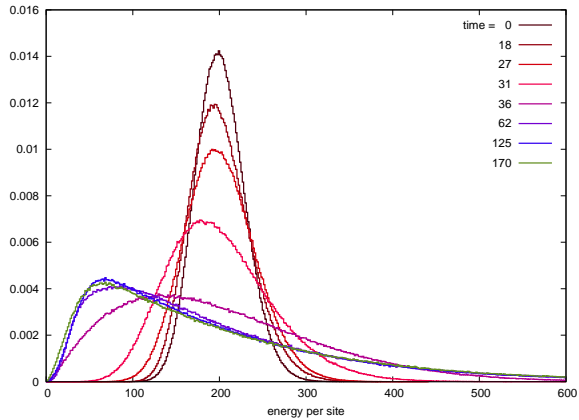}}}
\end{center}
\caption{\label{fig:Edist}Distribution of energy density at one
  lattice site, at various times in the evolution. The coupling
  constant is $g=0.5$.}
\end{figure}

\begin{figure}[htbp]
\begin{center}
\resizebox*{8cm}{!}{\rotatebox{-90}{\includegraphics{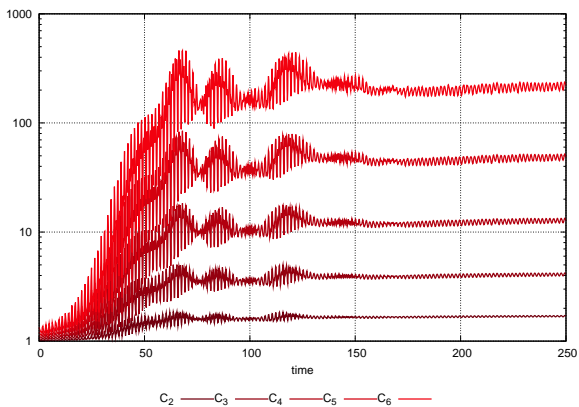}}}
\end{center}
\caption{\label{fig:Emoments}Normalized moments
  $\big<E^n\big>/\big<E\big>^n$ of the energy density distribution at 
  one lattice site, as a function of time. The coupling constant is
  $g=0.5$.}
\end{figure}

We show in figs.~\ref{fig:Edist} and \ref{fig:Emoments} results
from a study for the smallest subsystem one can conceive of on a
lattice--a single lattice site.
In fig.~\ref{fig:Edist}, we display histograms of the values of
the energy on one site\footnote{In lattice units, this is simply the
  value of $T^{00}$ at one given site.}, at various times in the
evolution. These curves are normalized so that their integral is
unity--they can be interpreted as probability
distributions for the value of the energy on one lattice site.  At
$t=0$, this distribution is very close to a Gaussian, centered on the
mean energy density in the system. The width of this Gaussian is
entirely determined by the Gaussian spectrum of fluctuations in
eq.~(\ref{eq:gaussian}). At early times, the distribution first
remains Gaussian-like, but tends to broaden with time. Around
$t\approx 30$ in lattice units, we observe a rapid change of shape of
this distribution--the peak of the distribution shifts to lower values
of the energy and the tail extends much further at large energy. Once
this dramatic change of shape has taken place, the evolution of the
distribution is rather slow and a stationary distribution is reached
at late times.

The evolution in the energy distribution can be explored further by
looking at its moments defined by
\begin{equation}
C_n\equiv \frac{\big<E^n\big>}{\big<E\big>^n}\; .
\end{equation}
Higher moments are very sensitive to changes in the shape of the
distribution, especially the appearance of an extended tail that
signals broader energy fluctuations.
We represent these moments as a function of time in
fig.~\ref{fig:Emoments}, up to $n=6$. They all start very close to 1
at $t=0$, which is the sign of a very narrow distribution with little
fluctuations. The rapid change of shape of the distribution around
$t\approx 30$ corresponds to a rapid increase of the moments. By
$t\approx 70$, the moments have reached nearly asymptotic values
modulo moderate residual oscillations.

It is interesting to compare the evolution of the energy distribution
at a single lattice site with the time evolution of the pressure in
fig.~\ref{fig:pressure1}. The initial rapid decrease of the pressure
oscillations is concomitant with the change of shape of the energy
distribution. The subsequent (slower) relaxation of the residual
oscillations of the pressure occurs after the energy has reached a
stationary distribution.

\section{Summary and Outlook}

We discussed in this paper a formalism which resums secular terms in a
weak coupling expansion of a scalar field theory with initial
conditions generated by strong sources. We showed that resummed
expressions, to all orders in perturbation theory, for inclusive
quantities could be expressed as an ensemble average of the
corresponding leading order classical quantities where the initial
classical field for each member of the ensemble is shifted by a
quantum fluctuation drawn from a Gaussian distribution. We showed that
this averaging caused the resummed pressure to relax to a single
valued relation with the energy density and interpreted this as
arising from the phase decoherence of individual classical
trajectories. We showed in a toy model that for an expanding system
our result leads to ideal hydrodynamical flow. We briefly addressed the
issue of thermalization--while our numerical results display features
similar to those of a canonical thermal ensemble, they differ slightly
in the particulars. A more systematic numerical study will likely be
able to shed further light on this important point. As noted in the
introduction, our system appears to satisfy Berry's conjecture which
has been argued to be an important requirement for the thermalization
of quantum systems. We plan to pursue this topic further in the future.

Finally, we note that to be fully relevant to heavy ion collisions,
our methods should be extended to gauge theories. We have shown
previously that the formalism outlined in section 2 here is also
applicable to a gauge theory~\cite{GelisLV2}. The spectrum of quantum
fluctuations~\cite{FukusGM1} is the essential ingredient here and work
on computing this quantity is well underway~\cite{DusliGSV1}.

\section*{Acknowledgements}
We would like to acknowledge useful discussions with Jean-Paul
Blaizot, Kenji Fukushima, Miklos Gyulassy, Tuomas Lappi, Larry
McLerran, Rob Pisarski, Andreas Schafer and Giorgio Torrieri. K.D's
and R.V's research was supported by DOE Contract No.
DE-AC02-98CH10886. F.G's work is supported in part by Agence Nationale
de la Recherche via the programme ANR-06-BLAN-0285-01. We thank the
Institute for Nuclear Theory at the University of Washington for its
hospitality. One of us (FG) would like to thank Brookhaven National
Laboratory as well as the Yukawa International Program for
Quark-Hadron Sciences at Yukawa Institute for Theoretical Physics
(Kyoto University) for partial support during the completion of this
work.

\appendix

\section{Lattice implementation}
\label{app:lattice}
In our numerical calculation, we discretize space in a $L^3$ cubic
lattice, while retaining time as a continuous variable. This means
that when solving the classical equation of motion for the field, the
timestep is freely adjustable in order to warranty a given accuracy.
In this appendix, we summarize the main aspects of this lattice
formulation.

\subsection{Discretization of space}
The field $\varphi(t,\x)$ (and its time derivatives) become a function of
a continuous time $t$ and of discrete indices $i,j,k$ that vary
between $0$ and $L-1$. We impose periodic spatial boundary conditions
to the fields. In order to keep the code simple, all the dimensionful
quantities are expressed in lattice units, which amounts to choosing a
lattice spacing $a=1$. Thus, the classical equation of motion for the
field becomes:
\begin{eqnarray}
\ddot{\varphi}_{ijk}(t)
&=&
\varphi_{i+1 jk}+\varphi_{i-1 jk}
+
\varphi_{ij+1k}+\varphi_{ij-1 k}
+
\varphi_{ijk+1}+\varphi_{ijk-1}
-6\varphi_{ijk}
\nonumber\\
&&
-\frac{g^2}{6}\varphi_{ijk}^3(t)+J_{ijk}(t)\; .
\label{eq:EOM-disc}
\end{eqnarray}
(The terms on the right hand side of the first line correspond to the 
discretized version of the Laplacian of the field.)

In our discretized description, there is an exactly conserved energy
at $x^0>0$.  To see that, multiply the previous equation by
$\dot\varphi_{ijk}$ and sum over all the lattice sites,
\begin{eqnarray}
&&
\frac{d}{dt}\sum_{ijk}\Big[\frac{1}{2}\dot\varphi_{ijk}^2
+\frac{1}{2}(\varphi_{i+1jk}\!-\!\varphi_{ijk})^2
+\frac{1}{2}(\varphi_{ij+1k}\!-\!\varphi_{ijk})^2
+\frac{1}{2}(\varphi_{ijk+1}\!-\!\varphi_{ijk})^2
\nonumber\\
&&\qquad\qquad\qquad+\frac{g^2}{4!}\varphi^4_{ijk}
\Big]
=
\sum_{ijk}J_{ijk}\dot\varphi_{ijk}\; .
\end{eqnarray}
The left hand side of this equation, which is nothing but the time
derivative of the total lattice energy, is zero when the source $J$ is
turned off. But note that for this to work, we had to use a
non-symmetric form of the discrete derivative in the definition of the
kinetic energy. Taking the seemingly more natural
$\frac{1}{2}(\varphi_{i+1jk}-\varphi_{i-1jk})$ instead of
$\varphi_{i+1jk}-\varphi_{ijk}$ would lead to an energy which is not exactly
conserved on the lattice (though violations of energy conservation
would be of higher order in the lattice spacing, it is preferable to
use a lattice definition that makes it exactly conserved for any
lattice spacing).

\subsection{Small field fluctuations}
To obtain the spectrum of fluctuations for the initial
condition of the classical field, we must also consider the time
evolution from $t=-\infty$ to $t=0$ of small perturbations to the
classical field that behave like plane waves in the remote past.  On
the lattice, free plane waves are labelled by three integers $l,m,n$
and are of the form:
\begin{equation}
  a^{(\pm lmn)}_{ijk}(t)
  \equiv
  e^{\pm i E_{lmn}t}\,e^{\mp\frac{2i\pi}{L} (il+jm+kn)}\; ,
\label{eq:freewave}
\end{equation}
where the energy of the mode $lmn$ is given by
\begin{equation}
E_{lmn}
=
\Big[
2\left(
3
-\cos\left(\frac{2\pi l}{L}\right)
-\cos\left(\frac{2\pi m}{L}\right)
-\cos\left(\frac{2\pi n}{L}\right)
\right)
\Big]^{1/2}\; .
\end{equation}
These plane waves obey the lattice Klein-Gordon equation,
\begin{equation}
\ddot{a}_{ijk}=
a_{i+1 jk}+a_{i-1 jk}
+
a_{ij+1k}+a_{ij-1 k}
+
a_{ijk+1}+a_{ijk-1}
-6a_{ijk}\; .
\end{equation}
The lattice version of the fluctuations $a_{\pm\k}(x)$ that enter in
eqs.~(\ref{eq:gaussian}) are fluctuations $a^{(\pm lmn)}_{ijk}(t)$ that obey
\begin{eqnarray}
&&
\ddot{a}_{ijk}+\frac{g^2}{2}\varphi^2_{ijk}\,a_{ijk}=
\nonumber\\
&&\qquad=
a_{i+1 jk}+a_{i-1 jk}
+
a_{ij+1k}+a_{ij-1 k}
+
a_{ijk+1}+a_{ijk-1}
-6a_{ijk}\; ,
\label{eq:fluct-evol}
\end{eqnarray}
and behave like the free waves of eq.~(\ref{eq:freewave}) when $t\to
-\infty$.

\subsection{Sampling of the Gaussian fluctuations}
Once the fluctuations that obey this equation and initial conditions
are known, the discrete version of eq.~(\ref{eq:gaussian}) reads
\begin{equation}
\left<a_{ijk}a_{i^\prime j^\prime k^\prime}\right>
=
\frac{1}{L^3}
\sum_{lmn}\frac{1}{2E_{lmn}}\;
a^{(+ lmn)}_{ijk}(0)\,
a^{(- lmn)}_{i^\prime j^\prime k^\prime}(0)\; ,
\label{eq:gaussian1}
\end{equation}
and similar formulas for the correlators involving the time
derivatives. Generating Gaussian fluctuations that have a given
correlation function in general requires to diagonalize the 2-point
correlator. On the lattice, this means that one should diagonalize an
$L^3\times L^3$ matrix, a fairly time consuming step even for a
reasonably sized lattice.

In the case of the correlation function (\ref{eq:gaussian1}), it turns
out that we can avoid this diagonalization. First of all, let us
decompose a fluctuation $a_{ijk}$ on the basis formed by the $a^{(\pm
  lmn)}_{ijk}$,
\begin{equation}
a_{ijk}
\equiv
\frac{1}{L^3}
\sum_{lmn}
\frac{1}{2E_{lmn}}\;
\Big[
\alpha_{lmn}\,a^{(+ lmn)}_{ijk}
+
\beta_{lmn}\,a^{(- lmn)}_{ijk} 
\Big]\; .
\label{eq:decomp}
\end{equation}
Since we want to obtain a real field, we must have
$\beta_{lmn}=\alpha_{lmn}^*$. Since there is a linear relation between
$a_{ijk}$ and the coefficients $\alpha_{lmn},\beta_{lmn}$, it is
obvious that these coefficients should also be Gaussian distributed.
A little guesswork indicates that in order to obtain
eq.~(\ref{eq:gaussian1}), one needs
\begin{eqnarray}
&&
\big<\alpha_{lmn}\alpha_{l^\prime m^\prime n^\prime}\big> = 
\big<\beta_{lmn}\beta_{l^\prime m^\prime n^\prime}\big> = 0\; ,
\nonumber\\
&&
\big<\alpha_{lmn}\beta_{l^\prime m^\prime n^\prime}\big> =
E_{lmn}\,L^3\,\delta_{ll^\prime}\delta_{mm^\prime}\delta_{nn^\prime}\; .
\end{eqnarray}
Equivalently, this can be translated into correlators for the real and
imaginary parts of $\alpha_{lmn}$:
\begin{eqnarray}
&&
\big<{\rm Re}\,(\alpha_{lmn}){\rm Re}\,(\alpha_{l^\prime m^\prime n^\prime})\big>=
\big<{\rm Im}\,(\alpha_{lmn}){\rm Im}\,(\alpha_{l^\prime m^\prime n^\prime})\big>
= \frac{E_{lmn}}{2}\,L^3\,\delta_{ll^\prime}\delta_{mm^\prime}\delta_{nn^\prime}
\nonumber\\
&&
\big<{\rm Re}\,(\alpha_{lmn}){\rm Im}\,(\alpha_{l^\prime m^\prime n^\prime})\big>=0
\; .
\end{eqnarray}
In words, the real and imaginary parts of the coefficient
$\alpha_{lmn}$ are independent Gaussian random variables. This leads
to a straightforward algorithm for generating the correct Gaussian
fluctuations of the initial conditions of the classical field:
\begin{itemize}
\item[{\bf i.}] From $t=-\infty$ (in practice, some large negative
  time) to $t=0$, solve the classical equation of motion
  (\ref{eq:EOM-disc}) for the classical field $\varphi$ given a source
  $J$,
\item[{\bf ii.}] For each Fourier mode $lmn$, solve the equation of
  motion (\ref{eq:fluct-evol}) for the small fluctuation $a^{(+lmn)}$,
  until $t=0$,
\item[{\bf iii.}] For each mode $lmn$, generate two random numbers
  ${\rm Re}\,(\alpha_{lmn})$, ${\rm Im}\,(\alpha_{lmn})$ drawn from
  a Gaussian distribution of variance $E_{lmn}L^3/2$,
\item[{\bf iv.}] Construct a field fluctuation $a,\dot{a}$ as
  \begin{eqnarray}
    a_{ijk}
    &\equiv&
    \frac{1}{L^3}\sum_{lmn} \frac{1}{E_{lmn}}\;
    {\rm Re}\,\Big[\alpha_{lmn}\,a^{(+lmn)}_{ijk}(0)\Big]
    \; ,
    \nonumber\\
    \dot{a}_{ijk}
    &\equiv&
    \frac{1}{L^3}\sum_{lmn} \frac{1}{E_{lmn}}\;
    {\rm Re}\,\Big[\alpha_{lmn}\,\dot{a}^{(+lmn)}_{ijk}(0)\Big]
    \; ,
  \end{eqnarray}
  and superimpose it on to the field $\varphi_{ijk}(0),\dot\varphi_{ijk}(0)$ in
  order to obtain a fluctuating initial condition for the evolution at
  $t>0$. Repeat steps {\bf iii} and {\bf iv} for each configuration in
  the Monte-Carlo evaluation of eq.~(\ref{eq:sum1}) in order to obtain
  an ensemble of initial conditions for the classical field.
\item[{\bf v.}] For each initial condition constructed in this way,
  solve the classical equation of motion (\ref{eq:EOM-disc}) for $t>0$
  (but now with $J=0$). Compute the energy-momentum tensor
  of this classical field. Average it over the ensemble of initial conditions.
\end{itemize}

\subsection{Ultraviolet sector}
The summation over fluctuations of the initial conditions for the
classical field in eq.~(\ref{eq:sum1}), if performed without any
constraint on the momentum of the fluctuations one includes, leads to
ultraviolet divergences in the energy-momentum tensor.

For instance, if we impose a cutoff $\Lambda$ on the momentum of the
fluctuations included in eq.~(\ref{eq:sum1}), one can check that
generically the energy density will contain terms of the form
\begin{equation}
\epsilon = \frac{Q^4}{g^2}
\oplus Q^2\Lambda^2\oplus \Lambda^4\; ,
\end{equation}
where $Q$ is the physical scale introduced via the source $J$.
Eq.~(\ref{eq:sum1}) is not renormalizable in the usual sense, because
it results from a resummation that mixes diagrams with arbitrarily
high loop orders, and it should be seen as an effective formula that
only makes sense with an ultraviolet cutoff. In order to include all
the Fourier modes that are subject to instabilities, we must choose
the cutoff such that $Q\lesssim\Lambda$. On the other hand, the cutoff
should not be too large, otherwise the result of eq.~(\ref{eq:sum1})
will be cutoff dependent. It turns out that at weak coupling ($g\ll
1$), there is ample room to choose such a cutoff (large enough to
encompass the relevant physics and small enough to keep the
cutoff-dependent terms small). Indeed, it is sufficient to have
\begin{equation}
Q\lesssim \Lambda \ll \frac{Q}{\sqrt{g}}\; .
\end{equation}
This window of allowed $\Lambda$'s can be enlarged if we notice
that the $\Lambda^4$ terms are pure vacuum contributions, that can
computed and subtracted once for all\footnote{This is done by running
  the same simulation with the source $J$ turned off, and by
  subtracting the corresponding result. We find that on a $12^3$
  lattice, the vacuum contribution to $T^{00}$ is almost independent
  of the coupling $g$, and equal to $T^{00}_{\rm vac}\approx 1.35$.
  The vacuum contribution to the pressure tensor is proportional to
  the identity, with a pressure equal to $\frac{1}{3}T^{00}_{\rm
    vac}$.}. After this subtraction has been performed, the condition
on $\Lambda$ becomes
\begin{equation}
Q\lesssim \Lambda \ll \frac{Q}{g}\; .
\end{equation}
Thus, taking $\Lambda$ of order $Q$ is a valid choice at small
coupling.

\subsection{Choice of the lattice cutoff}
We have seen in the previous subsection that a UV cutoff is necessary
in order to ensure the finiteness of eq.~(\ref{eq:sum1}).  One should
choose the cutoff so that all the modes up to the resonance band are
comprised below the cutoff. We have also seen that if the coupling
constant is small, then the dependence on such a cutoff is negligible
since it occurs only in terms that are subleading by one power of
$g^2$.

\begin{figure}[htbp]
\begin{center}
\resizebox*{8cm}{!}{\rotatebox{-90}{\includegraphics{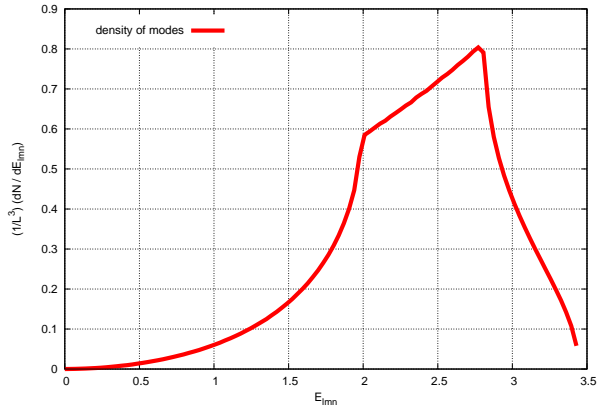}}}
\end{center}
\caption{\label{fig:density}Density of Fourier modes as a function of
  their lattice energy (in the limit $L\to \infty$).}
\end{figure}

Such a cutoff exists naturally on the lattice, as a consequence of the
discretization. In the figure \ref{fig:density}, we have represented
the density of lattice Fourier modes as a function of their energy
$E_{lmn}$.
This density falls abruptly when the energy reaches its maximal
allowed value $E=\sqrt{12}$ (in lattice units), and no Fourier mode with
a larger energy can exist on the lattice.

In lattice units ($a\equiv 1$), the interplay between the physical
scales and the lattice cutoff can be tuned by adjusting the parameter
$Q$ that sets the magnitude of the source $J$. The modes that are the
most important for the decoherence responsible for the relaxation
towards the equation of state are the modes $k\lesssim Q$. If we chose
a too large value of $Q$, the lattice cutoff will suppress modes that
are important for this relaxation process.  On the other hand, if $Q$
is too low, then the physically relevant modes fall in the region
where the density of lattice modes (see the figure \ref{fig:density})
is very small.  In this case, very few lattice modes are available to
represent the relevant physics. The optimal choice of $Q$ is to take
it so that the resonance band is located just before the fall off of
the mode density at large $E_{lmn}$. In this way, the physically
relevant modes sit in the region where the lattice mode density is the
largest.

\section{Linear stability analysis for $\phi^4$ field theory}
\label{app:stability}
\subsection{Instabilities of small perturbations}
The assumed $g^2$ suppression of the NLO correction relies on the fact
that the perturbations $\beta, a_{\pm\k}$ introduced in section
\ref{sec:NLO} have their ``natural'' order of magnitude
$(a_{\pm\k}\sim{\cal O}(1), \beta\sim{\cal O}(gQ))$. If their equation
of motion suffers from instabilities that amplify them, then the
previous estimate is incorrect and the NLO corrections may be as large
as the leading order contribution. It is therefore necessary to study
the stability of small perturbations to the classical field $\varphi$.

To keep the discussion simple, let us assume in this appendix that the
source $J$, and therefore also the classical field $\varphi$, do not
have any spatial dependence\footnote{The results we obtain here remain
  valid in the case their spatial gradients are small.}.  If
$a(x^0,\x)$ is a perturbation to the field $\varphi$, we can write its
evolution equation in a linearized form as long as it remains small
compared to the classical field $\varphi$ itself,
\begin{equation}
(\square + \frac{g^2}{2}\varphi^2) a =0\; .
\label{eq:lin-a0}
\end{equation}
This equation can be simplified by Fourier transforming the field
$a(x^0,\x)$ in the spatial variables\footnote{We use the same symbol
  for $a$ and for its Fourier transform to keep the notations light,
  as the context always allows one to distinguish the two.},
\begin{equation}
a(x^0,\x)
\equiv
\int\frac{d^3\k}{(2\pi)^3}\;a(x^0,\k)\;e^{i\k\cdot\x}\; ,
\end{equation}
so that
\begin{equation}
\ddot{a}+(\k^2+\frac{g^2}{2}\varphi^2) a=0\; .
\label{eq:lin-a}
\end{equation}
(The dot denotes a derivative with respect to time.) Given a pair of
solutions $a_1$ and $a_2$ of this equation, the Wronskian
$W\equiv\dot{a}_1a_2-a_1\dot{a}_2$ is independent of time.

When $\varphi$ depends only on time, it is a periodic function of time
at $x^0>0$.  Therefore, the coefficient of the term $a(x^0,\k)$ in
eq.~(\ref{eq:lin-a}) is also periodic in time, which may lead to
parametric resonance phenomena. For linear equations with periodic
coefficients, the stability analysis can be performed by finding the
``mapping at a period'', that evolves a pair of solutions $a_{1,2}$
from $x^0=0$ to $x^0=T$ (where $T$ is the period of the coefficients
in the equation):
\begin{equation}
\begin{pmatrix}
a_1(T,\k)       & a_2(T,\k)       \\
\dot{a}_1(T,\k) & \dot{a}_2(T,\k) \\
\end{pmatrix}
\equiv
M_\k
\begin{pmatrix}
a_1(0,\k)       & a_2(0,\k)       \\
\dot{a}_1(0,\k) & \dot{a}_2(0,\k) \\
\end{pmatrix}
\; .
\end{equation}
This mapping can be written as a multiplication by a matrix $M_\k$ thanks
to the fact that equation (\ref{eq:lin-a}) is linear. If the mapping
$M_\k$ is known, then after $n$ periods one has
\begin{equation}
\begin{pmatrix}
a_1(nT,\k)       & a_2(nT,\k)       \\
\dot{a}_1(nT,\k) & \dot{a}_2(nT,\k) \\
\end{pmatrix}
\equiv
M_\k^n
\begin{pmatrix}
a_1(0,\k)       & a_2(0,\k)       \\
\dot{a}_1(0,\k) & \dot{a}_2(0,\k) \\
\end{pmatrix}
\; ,
\end{equation}
and it is clear that the asymptotic behavior of the solutions
$a_{1,2}$ is determined by the eigenvalues $\lambda_{1,2}$ of the
matrix $M_\k$. 

\begin{figure}[htbp]
\begin{center}
\resizebox*{8cm}{!}{\rotatebox{-90}{\includegraphics{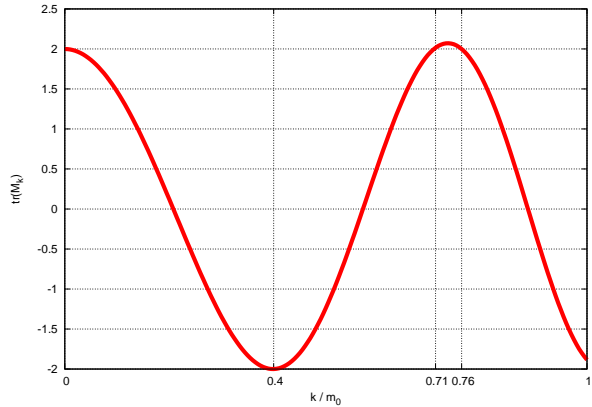}}}
\end{center}
\caption{\label{fig:trace}Trace of the monodromy matrix as a function
  of $km_0$ (we denote $m_0^2\equiv\frac{1}{2}g^2\varphi_0^2$).}
\end{figure}

From the conservation of the Wronskian, it is immediate to get
\begin{equation}
{\rm det}\,(M_\k)=\lambda_1\lambda_2=1\; .
\end{equation}
The two eigenvalues are thus mutually inverse, and we can
write the trace as:
\begin{equation}
{\rm tr}\,(M_\k)=
\lambda+\lambda^{-1}
\; ,
\end{equation}
where $\lambda$ is any of the two eigenvalues.  One has therefore
several cases:
\begin{itemize}
\item If ${\rm tr}\,(M_\k)>2$ (resp. ${\rm tr}\,(M_\k)<-2$), then
  $\lambda$ is real and greater than 1 (resp. smaller than -1). In
  this case, the solutions of eq.~(\ref{eq:lin-a}) generically diverge
  exponentially with time.
\item If ${\rm tr}\,(M_\k)=2$ (resp.  $-2$), then $\lambda=1$ (resp.
  -1). The two eigenvalues of $M_\k$ are in fact equal to $1$ (resp.
  -1). For $\lambda=1$, this implies that the matrix $M_\k$ is of the
  form
\begin{equation}
M_\k=P^{-1}
\begin{pmatrix}
1 & \alpha \\
0 & 1\\
\end{pmatrix}
P
\; ,
\quad
\mbox{and}\quad
M_\k^n
=
P^{-1}
\begin{pmatrix}
1 & n\alpha \\
0 & 1\\
\end{pmatrix}
P
\; .
\end{equation}
In this case, one of the solutions is $T$-periodic (and is therefore
stable), while the other solution of the basis diverges linearly with
time (unless $\alpha=0$ accidentally).
\item If $-2<{\rm tr}\,(M_\k)<2$, the eigenvalues $\lambda_{1,2}$ are
  complex and lie on the unit circle, $\lambda_{1,2}\equiv \exp(\pm
  i\theta)$ (they are mutual complex conjugates since the matrix
  $M_\k$ is real valued).  In this case, the small fluctuations are
  stable.
\end{itemize}
One can compute numerically the monodromy matrix $M_\k$ and its trace
as a function of $\k$. We have displayed the result in the figure
\ref{fig:trace}.
We see that the trace is equal to $2$ for a discrete set of modes,
including the zero mode\footnote{For $\k=0$, it is easy to check that
  $a_1(t)\equiv\dot\varphi(t)$ and $a_2(t)\equiv\dot\varphi(t)\int_0^t
  d\tau/\dot\varphi^2(\tau)$ are solutions of eq.~(\ref{eq:lin-a}).
  Given these two solutions, it is straightforward to get ${\rm
    tr}\,(M_0)=2$.}.  There is a band of modes\footnote{In section
  B.2, we show that the boundaries of this band are in fact
  $1/\sqrt{2}$ and $1/3^{1/4}$.} $0.71\le k/m_0\le 0.76$ in which the
trace is greater than 2, and where the fluctuations are exponentially
unstable.  Outside of these discrete modes and of the resonance band,
the trace is between -2 and 2, and the fluctuations are stable.
Besides these rigorous statements about the location of the unstable
modes, in practice the modes $\k$ for which $\big|{\rm
  tr}\,(M_\k)\big|<2$ but is close to 2 display a linearly growing
behavior for a rather long time, before they eventually decrease. For
instance, modes near the zero mode grow linearly for a time of order
$2\pi/k$ before they start decreasing.  Strictly speaking, these modes
are not unstable, but their growth at early times makes them important
for the transient dynamics of the system.

\subsection{Lyapunov exponent}
In the resonance band, the exponential instability of the solutions of
eq.~(\ref{eq:lin-a}) can be characterized by the Lyapunov exponent,
that one can define from the largest eigenvalue of $M_\k$ as follows,
\begin{equation}
\mu(\k,m_0)\equiv \frac{1}{T}\ln{\rm Max}\,\{\lambda_{1,2}\}\; ,
\end{equation}
where $m_0^2\equiv \frac{1}{2}g^2\varphi_0^2$.  Asymptotically, the
solutions $a(x^0,\k)$ of eq.~(\ref{eq:lin-a}) grow like
\begin{equation}
a(x^0,\k)\empile{\sim}\over{x^0\to+\infty} e^{\mu(\k,m_0)x^0}\; .
\end{equation}
The Lyapunov exponent can be obtained numerically from the trace of
$M_\k$, but in the case of the $\phi^4$ potential it can in fact be
derived analytically\footnote{The derivation we expose here is a
  particular case of techniques developed in \cite{GreenKLS1}.}.
Consider the equation
\begin{equation}
\ddot{a}+(\k^2+m^2(t))a=0\; ,
\label{eq:ee}
\end{equation}
where
\begin{equation}
\ddot{m}+\frac{1}{3}m^3=0\; .
\end{equation}
(In other words, $m(t)\equiv g\varphi(t)/\sqrt{2}$. The equation for
$m(t)$ is a consequence of the equation for $\varphi(t)$.)\footnote{The
  value of $g$ is not important per se.  Only the combination $m_0\sim
  g\varphi_0$ matters for the Lyapunov exponent.  Moreover, if
  $\mu(\k,m_0)$ is the Lyapunov exponent, then it scales as
\begin{equation}
  \forall\lambda\; ,\qquad  \mu(\lambda\k,\lambda m_0) = 
  \lambda\,\mu(\k,m_0)\; ,
  \label{eq:mu-scal}
\end{equation} due to the scale invariance of our model.} Let
us call $m_0$ the maximal amplitude of the oscillations of
$m(t)$, and introduce the new variable $z$ defined by
\begin{equation}
m^2=m_0^2 z\; .
\end{equation}
The time $t$ and the variable $z$ are related by
\begin{equation}
\frac{dz}{dt}=m_0 \sqrt{\frac{2}{3}z(1-z^2)}\; ,
\end{equation}
and we can rewrite eq.~(\ref{eq:ee}) as
\def\ppa{a^{\prime\prime}}
\def\pa{a^\prime}
\begin{equation}
2z(1-z^2)\ppa+(1-3z^2)\pa + 3(\kappa^2+z)a=0\; ,
\label{eq:e1} 
\end{equation}
where the prime denotes a derivative with respect to $z$ and
$\kappa\equiv k/m_0$. By this transformation, we have turned an
equation with oscillating coefficients into an equation with
polynomial coefficients. Given a pair $a_{1,2}$ of solutions of
eq.~(\ref{eq:e1}), one can show that the Wronskian is
\begin{equation}
W\equiv\pa_1 a_2-a_1\pa_2=\frac{w_0}{\sqrt{z(1-z^2)}}\; ,
\end{equation}
where $w_0$ is a constant.

\begin{figure}[htbp]
\begin{center}
\resizebox*{8cm}{!}{\rotatebox{-90}{\includegraphics{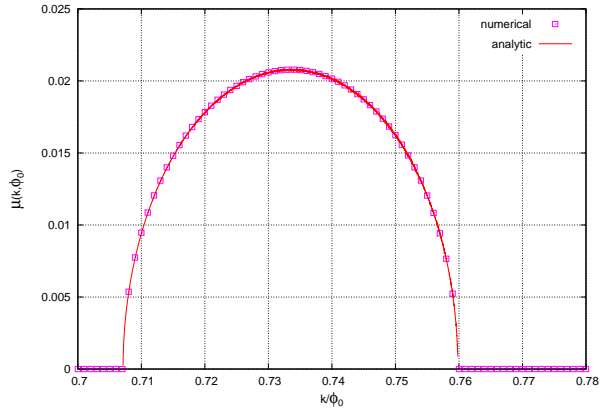}}}
\end{center}
\caption{\label{fig:muk-comp}Comparison between a numerical
  computation of the Lyapunov exponent, and its evaluation from
  eq.~(\ref{eq:lyapounov-final}).}
\end{figure}

Let us call now $M\equiv a_1 a_2$, with $a_{1,2}$ two solutions of
eq.~(\ref{eq:e1}) (possibly identical). A straightforward calculation
shows that $M$ obeys the following third order differential equation:
\begin{equation}
2 z(1-z^2)M^{\prime\prime\prime}+3(1-3z^2)M^{\prime\prime}
+6(z+2\kappa^2)M^\prime +6M=0\; .
\label{eq:e2}
\end{equation}
If $a_1$ and $a_2$ are two independent solutions of eq.~(\ref{eq:e1}),
then the three independent solutions of eq.~(\ref{eq:e2}) can be
thought of as $a_1^2$, $a_2^2$ and $a_1 a_2$. A remarkable property of
eq.~(\ref{eq:e2}) is that it admits a polynomial solution:
\begin{equation}
M(z)=z^2-2\kappa^2 z+4\kappa^4-1\; .
\end{equation}
If $\kappa$ is in the resonance band, where $a_1$ increases
exponentially while $a_2$ decreases exponentially, this polynomial
solution must be their product $M(z)=a_1(z)a_2(z)$. With the help of
the Wronskian, it is then easy to find
\begin{eqnarray}
a_1(z)&=&
\sqrt{|M(z)|}\,
\exp\left[+\frac{w_0}{2}\int^z \frac{dz}{M(z)\sqrt{z(1-z^2)}}\right]
\nonumber\\
a_2(z)&=&
\sqrt{|M(z)|}\,
\exp\left[-\frac{w_0}{2}\int^z \frac{dz}{M(z)\sqrt{z(1-z^2)}}\right]
\; .
\end{eqnarray}
In order to determine the constant $w_0$, one must insert these
solutions into eq.~(\ref{eq:e1}). This leads to
\begin{equation}
w_0=4\sqrt{6\kappa^2\left(\frac{1}{3}-\kappa^4\right)
\left(\kappa^4-\frac{1}{4}\right)}\; .
\end{equation}
The resonance band corresponds to the values of $\kappa$ such that the
argument of the square root is positive (otherwise $w_0$ would be
imaginary and one would have an oscillating solution instead of an
exponentially growing solution). Thus, the instability domain is
\begin{equation}
\frac{1}{\sqrt{2}}\le\kappa\le \frac{1}{3^{1/4}}\; .
\label{eq:band}
\end{equation}
From the above solutions $a_1,a_2$, one can also determine their
growth during one period of oscillation of $m(t)$, from which one gets
the Lyapunov exponent. One obtains
\begin{equation}
\mu(\k,m_0)T
=
2 w_0 \int_0^1\frac{dz}{M(z)\sqrt{z(1-z^2)}}\; ,
\end{equation}
where $T$ is the period of the oscillations of $m(t)$:
\begin{equation}
T=\frac{4\sqrt{6}}{m_0}\int_0^1\frac{dz}{\sqrt{1-z^4}}\; .
\end{equation}
We finally get:
\begin{equation}
\mu(\k,m_0)
=
2m_0
\sqrt{\kappa^2\left(\frac{1}{3}-\kappa^4\right)
\left(\kappa^4-\frac{1}{4}\right)}
\frac{\int_0^1\frac{dz}{(z^2-2\kappa^2 z+4\kappa^4-1)\sqrt{z(1-z^2)}}}
{\int_0^1\frac{dz}{\sqrt{1-z^4}}}\; .
\label{eq:lyapounov-final}
\end{equation}
The integrals in this formula can be evaluated
numerically\footnote{The integral in the numerator must be handled as
  a Cauchy principal value, since its integrand has two poles in the
  interval $z\in[0,1]$.}, and one can compare this analytical result
to the direct numerical computation (performed by computing the trace
of $M_\k$). As illustrated in the figure \ref{fig:muk-comp}, the
agreement is perfect.

\subsection{Linear instability and time decoherence}
Note that a zero Lyapunov exponent does not mean that the solutions
are stable, it only implies that they are not exponentially unstable:
outside of the resonance band determined by the
inequality~(\ref{eq:band}), the solutions of eq.~(\ref{eq:lin-a}) may
exhibit a linear growth in time -- either indefinitely if ${\rm
  tr}\,(M_\k)=2$ or during a finite time for all the other modes.  In
the case of the zero mode, this linear behavior has a very simple
interpretation, because it is a direct consequence of the fact that
the $\phi^4$ potential leads to non-harmonic oscillations. Indeed, the
classical field $\varphi$ oscillates periodically in the potential
with a frequency $\omega$ that is proportional to the amplitude
$\varphi_0$ of the oscillations\footnote{This result is specific to a
  potential that has only a $\phi^4$ term. Indeed, in four dimensions,
  such a field theory is scale invariant at the classical level, and
  its only dimensionful parameter is the amplitude of the oscillations
  of $\varphi$ (set by the external source $J$ that drives the system
  at $x^0<0$). For other potentials, the oscillation frequency
  $\omega$ is in general some complicated function of the amplitude
  $\varphi_0$ and of the coupling constants present in the
  potential.}: a classical field that oscillates freely in a $\phi^4$
potential can be written as
\begin{equation}
\varphi(t) = \varphi_0 f(\varphi_0 t)\; ,
\label{eq:scaling}
\end{equation}
where $f$ is a periodic function of amplitude unity.  Let us now add a
small perturbation $a$ to this classical field, so that its amplitude
is now $\varphi_0+a_0$. The perturbed oscillations are given by:
\begin{equation}
\psi(t)=(\varphi_0+a_0) f((\varphi_0+a_0)t)\; .
\end{equation}
If we assume that $a_0\ll \varphi_0$ and we linearize in the
perturbation, we have
\begin{equation}
\psi(t)=\varphi(t)
+ a_0 f(\varphi_0 t)
+ \varphi_0\, a_0 t\, f^\prime(\varphi_0 t)
+ {\cal O}(a_0^2)\; .
\end{equation}
The fact that the frequency depends on the amplitude produced a term
that grows linearly with time for the linearized perturbation. This
result is ubiquitous for any non-harmonic potential, and not specific
to $\phi^4$. This is the reason why a linear instability is the
generic behavior of the solutions of eq.~(\ref{eq:lin-a}). Due to its
origin, this linear instability is also closely related to another
property: two classical solutions that at $x^0=0$ differ only by a
small perturbation $a_0$ will eventually become completely incoherent
since their phase has shifted by $2\pi$ after a time of order
\begin{equation}
t_{\rm decoherence}\sim \frac{2\pi}{a_0}\; .
\end{equation}
In the case of perturbations that have a non-zero momentum $\k$, the
scaling law (\ref{eq:scaling}) is not valid if $g\varphi_0 \lesssim
|\k|$. For these high $\k$ modes, the oscillations are almost harmonic
and thus independent of their amplitude. This means that the linear
growth becomes weaker and weaker as $|\k|$ increases, and practically
irrelevant for Fourier modes above $g\varphi_0$.


\end{document}